\documentclass[acmsmall]{acmart}

\AtBeginDocument{%
  \providecommand\BibTeX{{%
    \normalfont B\kern-0.5em{\scshape i\kern-0.25em b}\kern-0.8em\TeX}}}

\setcopyright{cc}
\setcctype{by}
\acmJournal{PACMHCI}
\acmYear{2025} \acmVolume{9} \acmNumber{7} \acmArticle{CSCW413} \acmMonth{11} \acmPrice{}\acmDOI{10.1145/3757594}

\usepackage{xcolor}
\usepackage{color}
\usepackage{multirow}
\usepackage{makecell}
\usepackage{subfig}
\usepackage{ragged2e}
\usepackage{booktabs}
\usepackage{tabularx}
\definecolor{color1}{HTML}{40399F}
\definecolor{color2}{HTML}{1C4955}
\definecolor{color3}{HTML}{2973B8}
\definecolor{color4}{HTML}{D76364}
\definecolor{color5}{HTML}{4190A1}

\begin{document}

\title[Agency in Human-AI Co-creation: An HCI Perspective]{Exploring Collaboration Patterns and Strategies in Human-AI Co-creation through the Lens of Agency: A Scoping Review of the Top-tier HCI Literature}


\author{Shuning Zhang}
\orcid{0000-0002-4145-117X}
\email{zsn23@mails.tsinghua.edu.cn}
\affiliation{%
  \institution{Tsinghua University}
  \city{Beijing}
  \country{China}
}

\author{Hui Wang }
\orcid{0009-0003-1477-3338}
\email{hw1361376751@gmail.com}
\affiliation{%
  \institution{University of Duisburg Essen}
  \city{Essen}
  \country{Germany}
}

\author{Xin Yi}
\authornote{Corresponding author.}
\orcid{0000-0001-8041-7962}
\email{yixin@tsinghua.edu.cn}
\affiliation{
    \institution{Tsinghua University}
    \city{Beijing}
    \country{China}
}


\begin{abstract}

    As Artificial Intelligence (AI) increasingly becomes an active collaborator in co-creation, understanding the distribution and dynamic of agency is paramount. The Human-Computer Interaction (HCI) perspective is crucial for this analysis, as it uniquely reveals the interaction dynamics and specific control mechanisms that dictate how agency manifests in practice. Despite this importance, a systematic synthesis mapping agency configurations and control mechanisms within the HCI/CSCW literature is lacking. Addressing this gap, we reviewed 134 papers from top-tier HCI/CSCW venues (e.g., CHI, UIST, CSCW) over the past 20 years. This review yields four primary contributions: (1) an integrated theoretical framework structuring agency patterns, control mechanisms, and interaction contexts, (2) a comprehensive operational catalog of control mechanisms detailing how agency is implemented; (3) an actionable cross-context map linking agency configurations to diverse co-creative practices; and (4) grounded implications and guidance for future CSCW research and the design of co-creative systems, addressing aspects like trust and ethics. 
\end{abstract}


\received{October 2024}
\received[revised]{April 2025}
\received[accepted]{August 2025}

\begin{teaserfigure}
\includegraphics[width=\textwidth]{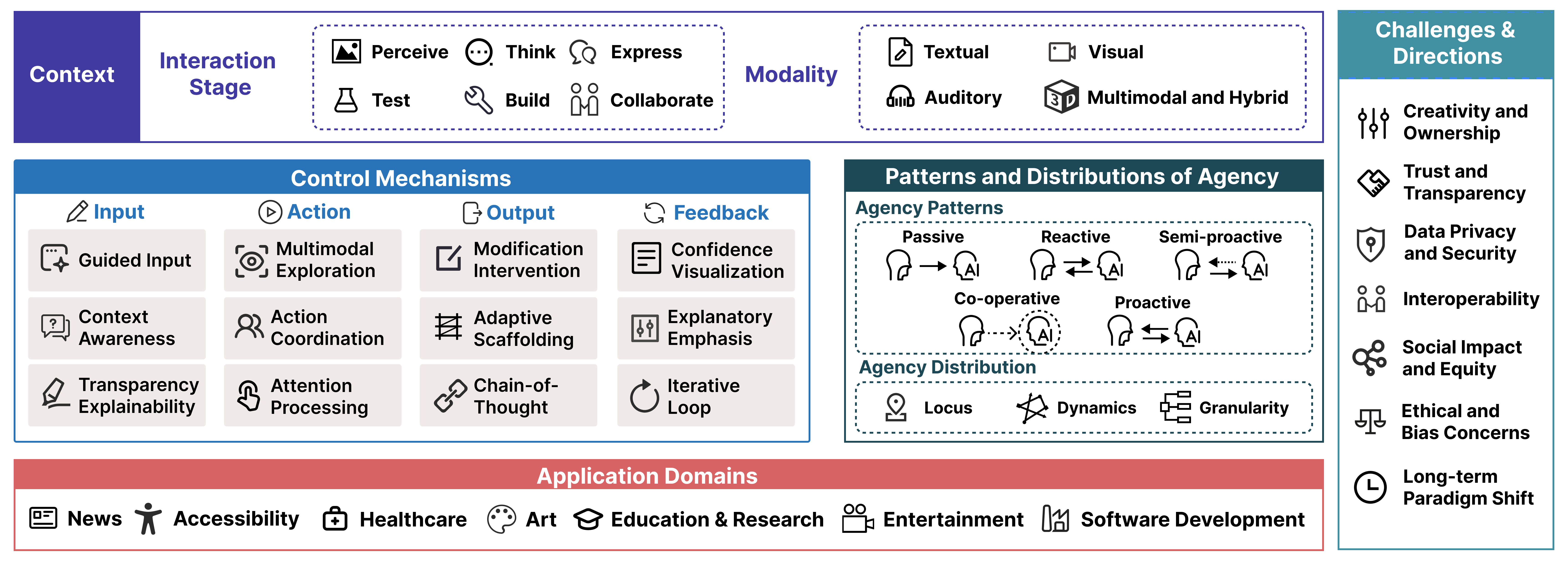}
\caption{The research framework for the scoping review of human-AI co-creation analyzed through the lens of agency. Reflecting on core user-centric design principles, the framework integrates the following dimensions: context (including interaction stage and modality), patterns and distributions of agency (including agency patterns and distribution), control mechanisms (including categories encompassing input, action, output and feedback), application domains, and challenges \& directions. Due to space constraints, we abbreviated the control mechanisms.}
\label{fig:frame}
\end{teaserfigure}

\maketitle

\section{Introduction}

Artificial Intelligence (AI) are increasingly transitioning from tools performing static tasks~\cite{eulerich2023can} to integral partners in collaborative processes like co-creation~\cite{zhang2023towards} and human-supervised autonomous systems~\cite{zhang2025actual}. This evolution marks a significant shift for Computer Supported Cooperative Work (CSCW), moving beyond traditional Human-Computer Interaction (HCI) paradigms~\cite{fitria2023artificial} to encompass scenarios where AI actively collaborates~\cite{wang2024promptcharm,pavlova2024analysis,fan2024contextcam}. As AI develops into a collaborative partner capable of proactive and potentially autonomous action~\cite{wu2021ai,moruzzi2024user,fu2020research}, it introduces fundamental challenges to the CSCW community in terms of the \textit{ distribution and negotiation of agency} (e.g. initiative, control), \textit{the operational mechanisms that enable this distribution}, and how these factors perform in various co-creative contexts and application domains~\cite{miernicki2021artificial, kahveci2023attribution}.

The concept of agency provides a critical lens for addressing the aforementioned complexities. Agency is understood here as the ability to act intentionally and exert control within a specific context~\cite{pacherie2007sense, limerick2014experience, aagerfalk2020artificial}, encompassing both human intentional action and the apparent agency perceived in machines. Although other valuable frameworks detail collaborative workflows~\cite{gao2024taxonomy,moruzzi2024user} or communication patterns~\cite{rezwana2022understanding}, they often underexplored AI's proactive capabilities. Consequently, they may not inherently provide the tool to systematically investigate the \textit{dynamic patterns of initiative} or the \textit{concrete operational mechanisms} through which control is enacted. The agency perspective is crucial, as it directly addresses these aspects, enabling a structured analysis of \textit{how} collaboration unfolds in terms of distributed action capabilities within various co-creative contexts.

Despite its importance, research applying the agency lens to human-AI co-creation is fragmented. Existing conceptualizations~\cite{wu2021ai,fu2020research} and frameworks~\cite{zhang2023towards,moruzzi2024user,zhou2024understanding} often lack granular detail on the specific operational control mechanisms needed to guide the design of practical systems. Although empirical studies~\cite{he2023exploring,guo2023rethinking,gao2024taxonomy} offer valuable situated insights, their focus on individual contexts limits the cross-context synthesis required to identify overarching patterns and design principles applicable across the co-creative landscape. Furthermore, investigations into mediating factors (e.g., interaction design~\cite{rezwana2023designing}, explainability~\cite{zhu2018explainable}) typically occur in isolation, limiting the structured mapping of how agency patterns, control mechanism, and contextual factors interrelate. This reveals a critical gap: \textbf{the lack of a comprehensive and systematic overview mapping the interplay between agency patterns, the control mechanisms implementing them, and the co-creative contexts (modalities, domains) where they are employed.}

To address this critical gap--particularly the need for a systematic synthesis of agency patterns and the operational control mechanisms implementing them across diverse contexts, this paper presents a scoping review. We focus specifically on research published in prominent HCI venues (e.g., CHI, UIST, CSCW) over the past 20 years, as this body of literature provides a concentrated documentation of the interaction designs and implemented control strategies relevant to operationalizing agency in co-creation systems. Our analysis covers 134 papers, employing the methodology detailed in Section 3 and proposing a novel framework (Figure~\ref{fig:frame}) to structure our investigation. The structure of the framework is informed by considerations of both system-level organization (integrating context, control, and application)~\cite{hu2025vision} and user-centric design principles~\cite{card2018psychology}.

Applying our analytical framework, the analysis yields specific insights into how the agency is operationalized in practice. We first elaborate on the context dimension (Section 4), identifying key situational factors from the reviewed literature. We then identify and classify the principal forms of AI agency currently employed in co-creative systems, revealing a spectrum from passive tools to cooperative partners (Section 5). We systematically catalog the concrete control mechanisms documented across interaction stages (input, action, output, feedback) used to manage agency distribution (Section 6). Furthermore, our synthesis reveals recurring associations linking specific control strategies (e.g. transparency, iterative feedback loops) and agency patterns with particular application domains and reported collaboration dynamics (Section 7). These findings directly inform our core contributions: the integrated analytical framework itself, the comprehensive catalog of operational control mechanisms, an actionable cross-context map of current practices, and grounded implications for the CSCW community (Sections 8-9).

\section{Related Work}

Agency is a foundational concept that is studied in diverse fields. Psychology clarifies its cognitive origins and perception mechanisms \cite{wegner2003mind,synofzik2008beyond}, sociology addresses complex questions of intentionality and responsibility \cite{list2011group,bratman2013shared}, while AI examines emerging machine capabilities and their perceptual effects \cite{sharma2024investigating,dai2024position}. These diverse inquiries collectively shape our understanding of human action, machine behavior, and their societal implications. Based on this broad context, our review adopts an HCI/CSCW lens. Here, AI's growing capabilities drive human-AI partnerships, bringing agency, control, and autonomy into the CSCW community's focus \cite{adenuga2024agency,cheon2021human,liao2019human}. We synthesize literature on agency allocation, the distinction between agency and control, and examine co-creation frameworks relevant to this scope.


\subsection{Agency and Its Distribution in Human-AI Collaboration}

Agency, central to HCI \cite{shneiderman2010designing}, yet often ambiguously defined \cite{bennett2023does}, encompasses both human and machine agency. Humans exhibit intentional agency, machines display apparent agency, the perception of thinking and acting capacities that machines seem to possess during interaction~\cite{takayama2015telepresence, brandtzaeg2023good}. The interplay between these forms, particularly in the closely coupled systems characteristic of human-AI collaboration \cite{cornelio2022sense, mueller2020next}, requires understanding the dynamic distribution of the agency between partners \cite{bennett2023does}.

With regard to this distribution, previous work explores both conceptual models and empirical studies. Conceptual frameworks, like multi-level models~\cite{zhang2023towards} or user-centered approaches~\cite{moruzzi2024user}, provide high-level structures but often lack granular detail on specific operational control mechanisms for practical design across diverse contexts~\cite{wu2021ai,fu2020research,zhou2024understanding}. For instance, Zhang et al.~\cite{zhang2023towards} did not categorize specific control mechanisms. While proposing dimensions for user configuration, Moruzzi et al.~\cite{moruzzi2024user} did not systematically synthesize the spectrum of control mechanisms or embed these in various contexts reported in the literature. Similarly, Zhou et al.~\cite{zhou2024understanding} proposed nonlinear co-design frameworks, but did not focus on synthesizing agency patterns and control mechanisms. Holter et al.~\cite{holter2024deconstructing} proposed an adaptation model to ageny interactions, but did not provide a granular and operational analysis of specific control mechanisms. 

Furthermore, empirical studies, while offering valuable insights~\cite{he2023exploring,guo2023rethinking,gao2024taxonomy}, are often limited to a specific context, limiting generalizability between settings. Gao et al.~\cite{gao2024taxonomy} for instance, emphasized interaction patterns, leaving agency dynamics and control mechanisms less examined. Furthermore, mediating factors like interaction design~\cite{rezwana2023designing} or explainability~\cite{zhu2018explainable} are often analyzed separately, obscuring their integrated role in agency dynamics. \textbf{Consequently, a systematic synthesis mapping agency patterns to their operational control mechanisms in various co-creative contexts is needed. This review aims to provide this integration.}

\subsection{Distinguishing Agency and Control in Human-AI Collaboration}

Analyzing human-AI collaboration requires clearly distinguishing between agency and control, although these concepts are often conflated~\cite{bennett2023does,zacarias2012human}. Agency primarily concerns intentional initiation, while control refers to the operational means of executing those intentions~\cite{pacherie2007sense,limerick2014experience}. This differentiation is crucial when interacting with AI partners capable of autonomous action~\cite{aagerfalk2020artificial}, where high-level user intent (agency) must interface with low-level system manipulation (control)~\cite{wang2024promptcharm}. The relevant theoretical work offers different perspectives~\cite{aagerfalk2020artificial, bennett2023does}. Specifically, Ågerfalk et al.~\cite{aagerfalk2020artificial} conceptualized AI as `a digital agency' capable of autonomous action (control) on behalf of humans (agency) and used context, communication and practice as key factors for analyzing this dynamic. Bennett et al.~\cite{bennett2023does} synthesized dimensions from the HCI literature on agency and autonomy (often used interchangeably with control). These dimensions include distinctions like self-causality versus identity, experience versus material reality, differing time-scales, and independence versus interdependence. While insightful, these abstract dimensions do not readily map to concrete control mechanisms. This review clarifies how the agency operates through specific controls by identifying and categorizing mechanisms from the literature.

\subsection{Human AI Co-Creation Framework}

Prior research offers foundational principles, conceptual models, and user-centric views relevant to human-AI co-creation, yet lacks a systematic synthesis detailing the interplay between agency patterns and the operational control mechanisms. Early foundational work established crucial interaction principles. Horvitz et al.~\cite{horvitz1999principles} defined mixed initiative principles that optimize interaction based on utility and uncertainty, while Amershi et al.~\cite{amershi2019guidelines} provided validated general human-AI interaction guidelines. While pioneering, this foundational work did not specify how the agency is operationally distributed through control mechanisms.

Conceptual models later structured the co-creation process. Wu et al.~\cite{wu2021ai} conceived AI roles (e.g. collaborator, tool) throughout creative stages, and Zhang et al.~\cite{zhang2023towards} proposed a three-level structure for AI involvement. Others specifically mapped text generation stages and controls~\cite{cheng2022mapping} or co-writing patterns~\cite{ding2023mapping}. These structural views offer valuable abstractions but lack detailed mapping of the concrete control mechanisms implementing agency variations, particularly across diverse, non-textual modalities.

Other studies adopted user-centered or context-specific lenses. Gmeiner et al.~\cite{gmeiner2023exploring}  highlighted designers' struggles to understand outputs and communicate goals when learning specific AI manufacturing tools. Kim et al.~\cite{kim2023one} classified ten AI roles prevalent in daily life and compared the perceptions of laypeople. Moruzzi et al.~\cite{moruzzi2024user} synthesized a user-centered framework that structures interactions through guidance, configuration, and dynamics. Although these studies illuminate user experiences and situated difficulties, they do not systematically map the underlying control mechanisms or the resulting agency distributions across different human-AI co-creative domains.

\section{Scope, Definitions and Methodology}

This paper presents a scoping review of human-AI co-creation, examining collaboration dynamics through the lens of agency. Our central aim is to understand how perceived human agency and machine agency influences the co-creative process and its results. This section first delineates the review's scope, and defines key terms (Section~\ref{sec:scope}). We then proposes the research framework (Section~\ref{sec:framework}) and the systematic systematic methodology employed for literature selection and analysis (Section~\ref{sec:literature}). Finally, we summarized our contributions  (Section~\ref{sec:contribution}). 

\subsection{Scope and Definitions}\label{sec:scope}

To ensure clarity and focus, this section defines the operational boundaries and key terminology used throughout this survey. These definitions are tailored to the objectives of our paper. 

\textbf{Human-AI Co-creation:} Generally recognized as a sub-field of human-AI collaboration~\cite{wu2021ai,gao2024taxonomy}, it lies at the intersection of Computational Creativity (CC) and HCI~\cite{moruzzi2024user}. This process often entails a dynamic partnership where humans and AI contribute unique strengths towards shared creative goals~\cite{prahalad2004co}, frequently involving what is termed ``AI creativity''. ``AI creativity'' emerged aside human creativity~\cite{turner2006artful}, marking its change towards a potential collaborator~\cite{boden1998creativity,miller2019artist}. Building on concepts like computational creativity~\cite{colton2012computational} and synergistic human-AI relationships~\cite{wu2021ai}, we adopt Wu et al.'s~\cite{wu2021ai} definition of ``AI creativity'' for this survey as: \textit{the capacity of AI systems to support and mutually facilitate human creative work by contributing distinct capabilities towards shared creative goals}. Integrating this aspect, we scope Human-AI Co-creation as: \textit{the activity where AI assists humans, often with a dynamic partnership and leveraging AI creativity, to complete creative works with each contributing unique efforts.} For this scope, we excluded: (1) works where AI performs only routine automation without creative contribution (e.g., basic spell checkers~\cite{MURRAY2024WHA}, simple data retrieval~\cite{10.5555/3495724.3496517}); (2) studies solely on human-human co-creation~\cite{10.1002/jocb.1519}; (3) AI assistance in purely analytical tasks lacking generative output~\cite{10.1145/3617362}.

\textbf{Agency (as an Analytical Lens):} To investigate the dynamics within co-creation, this review adopts agency as the central lens. Fundamentally, agency signifies the capacity to act or exert power~\cite{schlosser2015agency, liu2021ai}. Human agency is often linked to notions of intentionality, self-causality~\cite{bennett2023does}, and the experience of control over one's actions and their effects, sometimes conceptualized via `locus of control'~\cite{shneiderman2010designing}. Concurrently, advancements in AI have led to systems exhibiting increasingly autonomous behaviors~\cite{sundar2020rise}, giving rise to the concept of machine agency. Generally, this is understood as the capacity of machines to act autonomously and interact with users or environments~\cite{brandtzaeg2023good}. Despite the general concepts, analyzing agency within interactive systems requires a focused viewpoint. 

\textbf{The HCI/CSCW Perspective on Agency:} The fundamentally interactive nature of human-AI co-creation necessitates an analytical approach grounded in HCI/CSCW, a discipline concentrating on the design, evaluation, and implementation of interactive computing systems for human use~\cite{card2018psychology}.  Distinct from philosophical inquiries into true intentionality~\cite{floridi2014fourth} or AI research concentrated on autonomous capabilities~\cite{russell2016artificial}, our HCI/CSCW viewpoint specifically examines how agency, including both human agency and the agency attributed to machines, is \textit{manifested, perceived, negotiated, and managed through the design and use of the interactive system}~\cite{shneiderman2022human}. This perspective prioritizes user experience, the allocation and dynamics of control, system transparency, and the practical effectiveness of the human-AI partnership in forming a ``closely coupled system''~\cite{cornelio2022sense,mueller2020next}. Key concerns involve supporting users' sense of control and understanding how system behaviors are interpreted as agentic~\cite{bennett2023does}. 

\textbf{Agency and Control within HCI/CSCW:} Operating from this HCI/CSCW perspective, we delineate agency and control, two core definitions. Agency, within this context, primarily pertains to the intentional initiation of action and the subjective experience of directing outcomes toward specific goals~\cite{pacherie2007sense,limerick2014experience}. Crucially for HCI and CSCW communities, this includes not only human intentional action but also the apparent agency attributed to AI systems~\cite{takayama2015telepresence}. It is imperative to distinguish this higher-level sense of agency from Control. Control denotes the operational aspect, specifically the concrete mechanisms, commands, and manipulations employed by the user (or system) to execute intentions and influence system parameters during interaction~\cite{pacherie2007sense,wang2024promptcharm,aagerfalk2020artificial}. This distinction, often blurred but critical~\cite{bennett2023does,zacarias2012human}, allows for a multi-level analysis of co-creative interaction, examining both high-level goal-directed guidance (agency) and the specific means of execution (control)~\cite{bennett2023does}.

\textbf{Literature Scope and Justification:} To ensure the quality and relevance of the literature underpinning this review, we implement a focused scoping strategy. Our analysis concentrates primarily on research published within leading, peer-reviewed HCI/CSCW conferences, the principal forums where state-of-the-art HCI/CSCW research is presented, scrutinized, and debated. Our strategic focus is essential for engaging directly with methodologically sound studies and central theoretical advancements pertinent to human-AI co-creation and agency. Prioritizing these high-impact venues allows us to capture the core discourse and significant contributions within the field. \textbf{Crucially, this selective methodology guarantees that our synthesis is built upon a foundation of rigorously reviewed research shaping the HCI/CSCW understanding of human-AI partnership.}

\subsection{Framework}\label{sec:framework}

To systematically address the identified gap regarding the operationalization of agency across diverse human-AI co-creation contexts~\cite{miernicki2021artificial,kahveci2023attribution}, specifically the lack of a comprehensive overview linking agency patterns, control mechanisms and context factors, we developed the analytical framework in Figure~\ref{fig:frame}. This framework provides a structured lens to synthesize fragmented HCI literature~\cite{zhang2023towards, moruzzi2024user, zhou2024understanding, he2023exploring, gao2024taxonomy}. Its design integrates a multi-level, system-oriented perspective \cite{hu2025vision} with user-centric design principles \cite{card2018psychology} to analyze the agency landscape and address practical CSCW challenges. The framework integrates key dimensions reflecting a logical flow: the surrounding \textbf{context}, the core \textbf{patterns and distributions of agency}, the implementing \textbf{control mechanisms}, specific \textbf{application domains}, and overarching \textbf{challenges \& directions}, enabling a systematic mapping of how agency manifests, where, and how it is interactively achieved, and the associated difficulties.

Each dimension is critical for understanding the operationalization of agency. Because interaction fundamentally depends on its specific circumstances (e.g., task, environment) \cite{suchman1987plans}, analyzing \textbf{context} is essential for synthesis beyond isolated case studies. We thus examine interaction stage \cite{wu2021ai, zhang2023towards} to account for stage-specific agency patterns, and interaction modality \cite{shi2023hci}, as the channel shapes control possibilities. The \textbf{patterns and distributions of agency} dimension serves as the core theoretical lens, through which we examine AI initiative via agency patterns (e.g., passive to co-operative \cite{rammert2008action}) and authority management via agency distribution (locus, dynamics, granularity), extending prior conceptualizations \cite{holter2024deconstructing} to enable structured analysis. The \textbf{control mechanisms} dimension catalogues the concrete means through which agency is operationalized, offering actionable design insights for the CSCW community, with mechanisms (e.g., Guided Input, Iterative Feedback Loops \cite{hoque2024hallmark, liu2024coquest, weisz2022better}) organized within an Input-Action-Output-Feedback interaction cycle \cite{wiener2019cybernetics}. The \textbf{application domains} dimension grounds the analysis by mapping these findings onto specific fields (e.g., healthcare, software development \cite{wu2021ai}), thereby increasing the actionability of the insights. Finally, the \textbf{challenges \& directions} dimension synthesizes persistent issues (e.g., societal implications, ethics, long-term paradigm shift), guiding future research and responsible design.

\subsection{Methodology}\label{sec:literature}

We wanted to select conferences that are the most central and concentrated area where this type of interaction-centric HCI research is typically presented and debated. Thus, we conducted a scoping review on top-tier HCI conferences. Our methodology followed the principles outlined in the PRISMA-ScR (Preferred Reporting Items for Systematic Reviews and Meta-Analyses, Extended for Scoping Reviews) guidelines~\cite{tricco2018prisma}. The process involved structured literature searching and selection, followed by analysis and synthesis of the identified papers (detailed justifications and figures shown in Appendix~\ref{sec:prisma} and Figure~\ref{fig:prisma}).

\textbf{Literature Search and Selection}: Our primary search database was the ACM Digital Library, selected for its comprehensive coverage of premier HCI venues. Targeting publications up to August 2024, the search terms included combinations of (``co-creation'' OR ``co-writing'' OR ``co-drawing'' OR ``co-design'') AND (``agency''), applied to titles and full text. This strategy aimed to capture research at the intersection of HCI processes and agency manifestation. This resulted in 2799 records. After selecting HCI venues and scoping the time to recent 20 years, this resulted in 728 records.

\textbf{Screening} The initial search yielded 728 records. These underwent a screening process conducted by two researchers based on pre-defined criteria. We screened titles and abstracts for relevance to the core topic of human-AI co-creation and agency. We excluded records for two reasons: SC1. irrelevant topics: papers whose titles or abstracts clearly indicated a focus outside the scope of human-AI co-creation and agency (N=146 excluded). SC2. not full paper: records that were not peer-reviewed full papers (e.g., workshop summaries, posters) (N=46 excluded). A total of 192 records were excluded during this initial screening, leaving 536 papers eligible for full-text review.

\textbf{Eligibility} The remaining 536 papers were assessed for eligibility through full-text review. We applied three criteria: EC1. Relevance to Co-creation: papers not focused on collaborative or co-creative processes within an HCI context were excluded (N=185 excluded, e.g., algorithm optimization, autonomous navigation). EC2. Focus on concrete HCI contributions: papers discussing agency or co-creation at a conceptual, philosophical, ethical level without detailing or evaluating HCI techniques, interaction designs, or systems were excluded (n=95 excluded, e.g., high-level framework proposals). EC3. Focus on human-AI interaction: papers involved only human-human or AI-AI collaboration were excluded (N=122 excluded). This process resulted in 134 papers for detailed analysis.

\textbf{Analysis and Synthesis}: We used a hybrid thematic analysis approach~\cite{braun2006using}, a methodology well-suited for literature reviews requiring applications of existing theories and discovery of emergent patterns~\cite{spektor2023charting}. Two researchers jointly coded the final 134 papers, with intermittent discussions to solve disagreements. Our coding framework integrated both deductive and inductive dimensions: deductive coding applied established theoretical frameworks, while inductive coding identified and refined themes directly from data presented in the papers (details see Appendix~\ref{sec:coding}).

\textbf{Overview of Selected Papers}: The final corpus of 134 papers reflects diverse research approaches. As illustrated in Figure 2, qualitative research (N=62) and user-centric design (N=61) were the most common methods, followed by mixed-methods research (N=49). In terms of contribution types, design (N=98) and system implementation (N=66) were predominant, with significant contributions also coming from interviews (N=59). These papers were primarily from recent years (Figure 3), which reflected that the active role of AI propelled discussions of agency and co-creation. These papers are drawn from top-tier conferences with the following distribution: 108 CHI\footnote{The ACM Conference on Human Factors in Computing Systems} papers, 4 CSCW\footnote{The ACM Conference on Computer Supported Cooperative Work} papers, 5 DIS\footnote{The ACM Conference on Designing Interactive Systems} papers, 9 IUI\footnote{International Conference on Intelligent User Interfaces} papers and 8 UIST\footnote{ACM Symposium on User Interface Software and Technology} papers.

\begin{figure}[!htbp]
    \subfloat[Research methods.]{
        \includegraphics[width=0.46\textwidth]{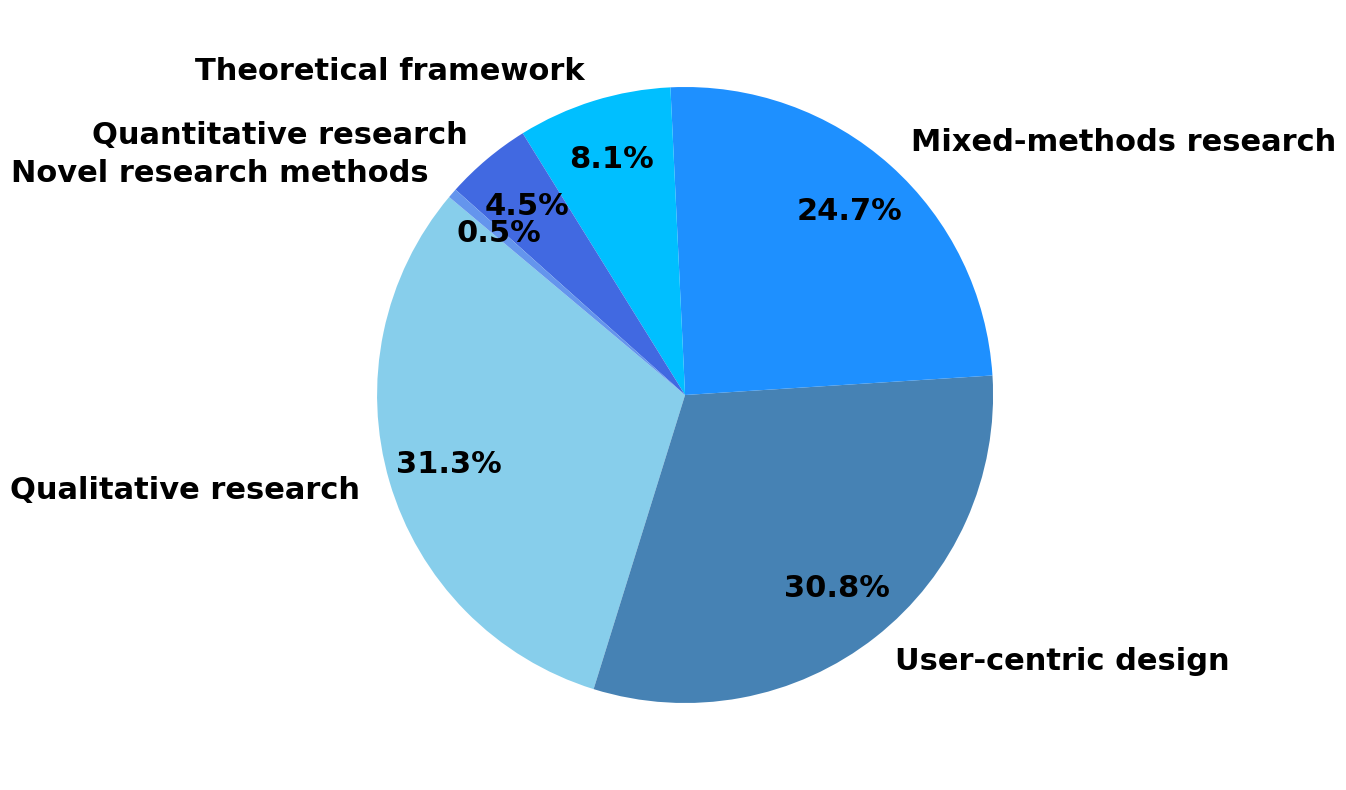}
        \label{fig:research_method}
    }
    \subfloat[Contribution types.]{
        \includegraphics[width=0.46\textwidth]{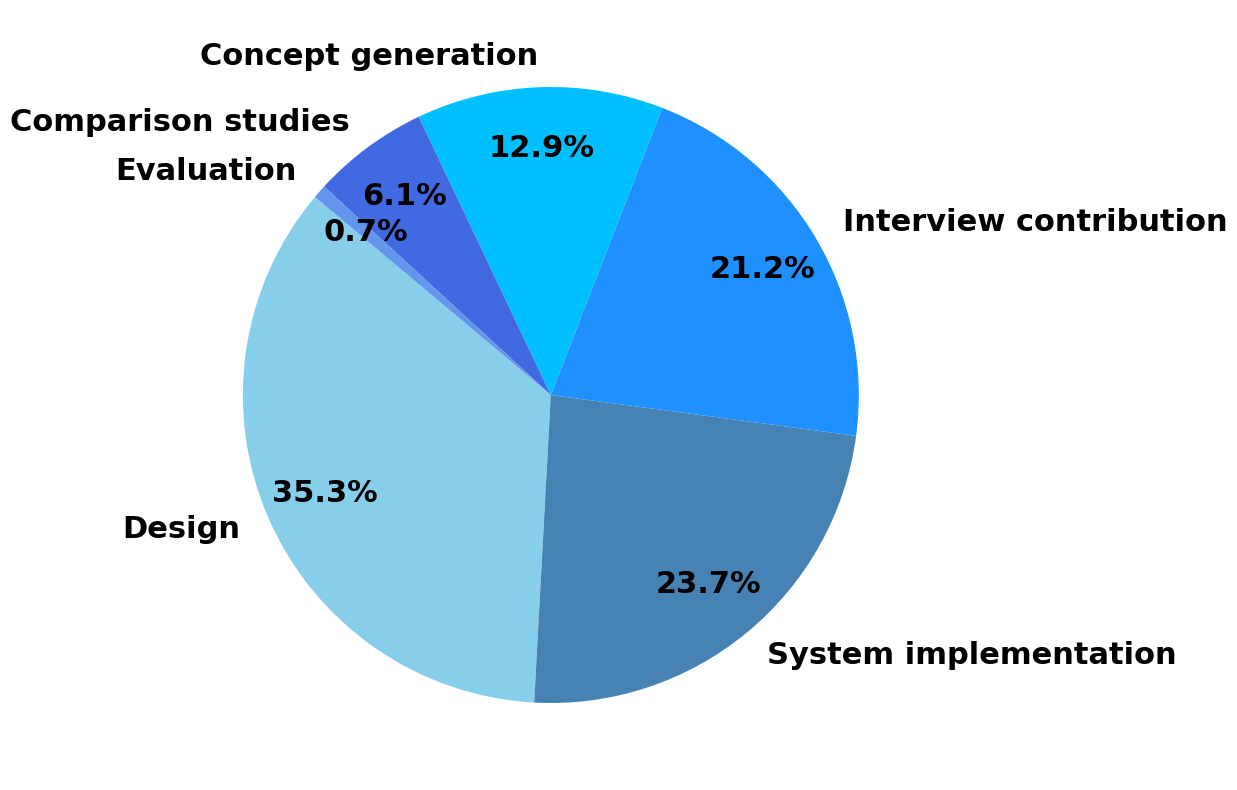}
        \label{fig:contribution_type}
    }
    \caption{The research methods and contribution types of the papers.}
    \label{fig:research_contribution}
\end{figure}

\begin{figure}  
    \centering
    \includegraphics[width=0.6\textwidth]{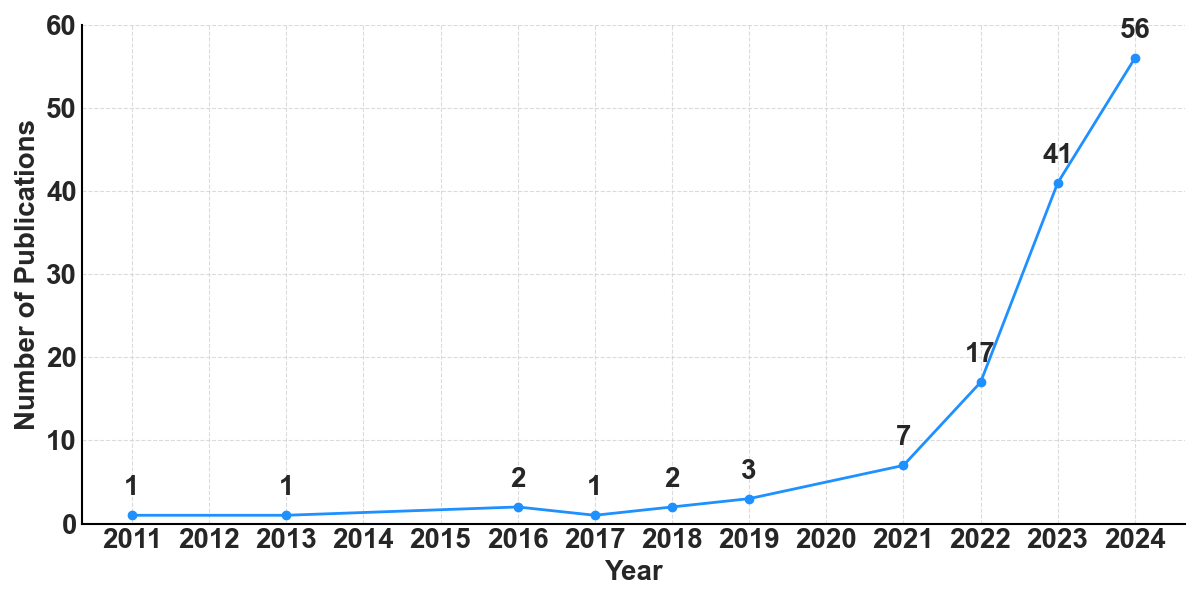}
    \caption{The publication numbers per year from our selected 134 papers.}
\end{figure}

\subsection{Contributions}\label{sec:contribution}

Our review adopts \textbf{agency as a central analytical lens} to offer distinct contributions, aimed at informing the CSCW community's understanding and design of human-AI collaboration:

\textbf{[Theoretical] An integrated framework for agency analysis:} We propose a multi-dimensional analytical framework (Figure 1) unifying agency patterns, operational control mechanisms, and interaction contexts, providing a necessary theoretical structure for analyzing human-AI co-creation.

\textbf{[Operational] Comprehensive catalog of control mechanisms:} We deliver a comprehensive categorization of concrete control mechanisms from HCI practice, detailing how agency is operationalized and informing system design.

\textbf{[Actionable] Cross-context map of co-creation practices:} Through extensive synthesis (134 papers in HCI venue), we construct an actionable map linking agency configurations (patterns, mechanisms) to co-creative contexts, offering a structured, cross-context overview that counters research fragmentation.

\textbf{[Implications] Grounded guidance for CSCW:} We derive specific design implications and future research directions from the synthesis, providing evidence-based material for deeper CSCW discussions on trust, ethics, and the design of future co-creative systems.

\section{Context}

\subsection{Interaction Stage}

There are many frameworks for classifying interaction stages. Although co-creation literature presents frameworks such as Guo et al.'s discover, define, develop, deliver model~\cite{guo2023rethinking}, their framework were refined to designers and may not reflect the broad practice in co-creation. Wu et al.~\cite{wu2021ai} proposed a process-oriented classification--perceive, think, express, collaborate, build, and test--which Zhang et al.~\cite{zhang2023towards} subsequently utilized for co-creation taxonomies. Recognizing that agency manifests dynamically throughout this interactive process, we adopt Wang et al.'s six-stage classification to systematically analyze the context of agency.

\textbf{\textcolor{color1}{Stage-1. Perceive:}} This initial stage involves both human and AI actors gathering and interpreting information to establish a shared understanding of the task context and goals.  The human formulates intent and provides initial input, which can range from textual prompts~\cite{fu2024text} and visual data~\cite{fan2024contextcam,verheijden2023collaborative} to audio~\cite{kamath2024sound,louie2020novice}, video~\cite{lyu2024preliminary,wang2024critical}, or embodied inputs like gestures~\cite{cremaschi2024steampunk,zheng2024charting,bircanin2021including}. The clarity and specificity of human input critically shape the AI's perception and subsequent actions, representing a primary locus of human agency in directing the process. Concurrently, the AI perceives user input and contextual data, activating its internal models. The effectiveness of this stage affects the AI's ability to accurately interpret diverse inputs and the human's skill in articulating goals, directly influencing human and AI agency.

\textbf{\textcolor{color1}{Stage-2. Think:}}
Following perception, this stage encompasses the internal cognitive or computational processes where both human and AI formulate ideas, strategies, and potential solutions. For the AI, this involves processing the perceived input using underlying algorithms for tasks such as natural language understanding~\cite{sharma2024generative,weisz2022better,dhillon2024shaping}, image recognition or generation~\cite{benjamin2023entoptic,fan2024contextcam,lin2024text}, data analysis~\cite{ayobi2023computational}, qualitative reasoning~\cite{gebreegziabher2023patat,tholander2023design}, or decision-making~\cite{berge2023designing}. Human agency is exercised through mental modeling, planning, and evaluating possibilities, often informed by prior knowledge and intuition. AI agency in this stage relates to its computational capabilities and algorithmic autonomy to generate novel or relevant intermediate representations based on its task interpretation. The alignment (or misalignment) between human and AI `thinking' critically impacts the collaborative trajectory.

\textbf{\textcolor{color1}{Stage-3. Express:}}
This stage involves articulating and externalizing internally generated ideas or results. The AI expresses its `thoughts' by generating outputs, manifested as text~\cite{sun2024metawriter,sharma2024generative,fu2024text}, images~\cite{fan2024contextcam,verheijden2023collaborative}, sounds~\cite{louie2020novice,sun2024understanding}, multimodal content~\cite{ayobi2023computational}, embodied actions~\cite{mirowski2023co,cremaschi2024steampunk}, or decisions~\cite{berge2023designing}. The clarity, relevance, and interpretability of the AI's expression are crucial for human comprehension and subsequent action, influencing the perceived usefulness and agency of the AI partner. Simultaneously, humans express their ideas, refinements, or directives, often through further input or manipulation of the AI's output. Human agency is prominent here in selecting, filtering, and communicating ideas, while AI agency is perceived through the form and content of its generated expressions.

\textbf{\textcolor{color1}{Stage-4. Collaborate:}}
The stage embodies the core interactive partnership, characterized by iterative exchanges aimed at refining ideas, resolving discrepancies, and jointly advancing the creative endeavor~\cite{rezwana2022understanding,holter2024deconstructing}. It often involves turn-taking and negotiation~\cite{ashktorab2021effects}, where humans critique or modify AI suggestions~\cite{yuan2022wordcraft,yang2022ai}, and the AI responds to human feedback, potentially asking clarifying questions~\cite{hodhod2016closing} or offers alternatives~\cite{mirowski2023co,wu2022ai}. Agency during collaboration becomes highly distributed and dynamically negotiated~\cite{holter2024deconstructing,lee2024and}. Effective collaboration requires mechanisms that foster mutual understanding and coordinated action. Transparency and explainability regarding AI actions~\cite{zhu2018explainable,hoque2024hallmark,wang2024farsight} are crucial for enabling humans to appropriately trust and engage with AI contributions~\cite{zhang2022get,weisz2022better}. Furthermore, well-designed communication protocols and interaction modalities~\cite{rezwana2023designing,mccormack2019silent} are vital for balancing human directorial control~\cite{fu2024text} with productive AI initiative~\cite{horvitz1999principles}, facilitating synergistic outcomes~\cite{weisz2022better,suh2021ai}. The perceived role of the AI (tool, teammate, or expert) significantly shapes these dynamics~\cite{zheng2024charting,kim2023one}.

\textbf{\textcolor{color1}{Stage-5. Build:}}
This stage translates collaboratively refined concepts into tangible artifacts or implemented solutions. It involves concrete construction, where humans might assemble physical components~\cite{bircanin2021including}, write or finalize code~\cite{weisz2022better,jonsson2022cracking}, render detailed designs~\cite{lawton2023drawing}, or structure narrative elements~\cite{osone2021buncho}. AI contributions generated in earlier stages often serve as foundational blocks, templates~\cite{jiang2022promptmaker}, or reference materials during building. Furthermore, AI can actively participate in this stage by automating specific construction tasks (e.g., code translation~\cite{weisz2022better}, component generation~\cite{hegemann2023computational}), simulating outcomes to guide refinement~\cite{kang2024design}, performing detailed modifications under human guidance~\cite{kamath2024sound,shi2020emog}, or assisting in the fabrication process itself~\cite{albaugh2023augmented,zhou2023beyond}. Human agency often pivots towards execution, integration, and fine-grained control~\cite{verheijden2023collaborative} during this stage, materializing the co-created vision. The manifestation of AI agency can vary significantly, ranging from a passive resource provider to an active co-constructor~\cite{vear2024jess+}, contingent upon the system's capabilities and the predefined collaborative workflow~\cite{zheng2024charting}.

\textbf{\textcolor{color1}{Stage-6. Test:}}
Finally, this stage encompasses evaluating the co-created output against initial goals or emergent criteria. Humans assess the artifact's quality, functionality, and alignment with their vision. This evaluation often involves providing feedback, which can loop back to earlier stages for iteration. Feedback modalities can range from explicit interface-based adjustments~\cite{kamath2024sound,zhang2022get} and natural language corrections~\cite{fan2024contextcam} to ratings~\cite{shaer2024ai} or embodied reactions~\cite{bauer2024musictraces}, reflecting nuanced qualitative assessments~\cite{shaer2024ai} or sometimes minimal explicit feedback~\cite{wan2024metamorpheus}. Human agency is paramount in judging the outcome and deciding on revisions, while AI agency might exist in automated testing or learning from evaluations to improve future performance. This stage closes the loop, fostering adaptation and learning for both partners.

\subsection{Interaction Modality}

Human-AI co-creation systems facilitate collaboration through diverse interaction modalities, shaping the nature of the partnership and the distribution of agency~\cite{shi2023hci}. While prior work has proposed categorizations based on the type of creative output or domain, such as text and language, visual arts, music/audio, and software development~\cite{JustThinkAI2024CoCreation}, or the specific data type involved like textual, 2D visual, layout, numerical, audio and 3D graphics~\cite{shi2023hci}, this section classifies interaction based on the primary channel through which humans and AI exchange information and exert influence during the collaborative process. We identify four principal modalities, acknowledging that practical applications often involve hybrid approaches.

\textbf{\textcolor{color1}{Modality-1. Textual Interaction:}} This modality centers on the collaborative manipulation of symbolic language. Humans and AI partners engage primarily through written text for generation, refinement, and analysis. Examples include AI assisting in generating or suggesting edits for code~\cite{wu2022ai,weisz2022better,choi2023creator}, where AI suggestions can reduce errors~\cite{weisz2023toward}, or enhancing academic writing~\cite{dhillon2024shaping,sun2024metawriter,park2023importance}, where appropriate scaffolding significantly improves outcomes, especially for less experienced users~\cite{dhillon2024shaping}. This modality also underpins collaborative ideation~\cite{sharma2024generative,zhang2024mathemyths}, story editing~\cite{yuan2022wordcraft}, and structured exploration of AI-generated responses to advance creative workflows~\cite{suh2024luminate}. The core interaction involves linguistic exchange, leveraging AI's capacity for processing and generating coherent and contextually relevant text.

\textbf{\textcolor{color1}{Modality-2. Visual Interaction:}} Collaboration in this modality revolves around the creation, modification, and interpretation of visual information, encompassing both static and dynamic forms. This includes co-drawing, where AI assists in generating sketches or suggesting palettes or visual artworks~\cite{de2024bias,almeda2024prompting,shi2020emog} or daily images~\cite{fan2024contextcam,lin2024text,lyu2024preliminary}, with text-to-image prompts notably facilitating verbal articulation in design ideation~\cite{lin2024text}. It also extends to video co-editing for applications like filmmaking~\cite{praveena2023exploring}, short video generation~\cite{wang2024critical}, and content moderation~\cite{lukoff2021design,choi2023creator}. Here, AI capabilities in scene detection, clip selection, or effect application augment human creative direction. Studies highlight factors influencing adoption in areas like robotic cinematography~\cite{praveena2023exploring}, the impact of algorithmic recommendations on user agency~\cite{lukoff2021design}, creator adaptation strategies~\cite{choi2023creator}, and the use of visual media for cultural heritage promotion~\cite{wang2024critical}. Reframing co-creative practices through post-human perspectives also informs this space~\cite{tholander2023design}.

\textbf{\textcolor{color1}{Modality-3. Auditory Interaction:}} This modality focuses on collaborative engagement through sound, including music creation, audio synthesis, and soundscape design. AI partners can generate musical ideas, manipulate sonic parameters, or assist in refining audio compositions. Examples include leveraging AI ambiguity creatively in sound design through generative Creative Support Tools (CSTs)~\cite{kamath2024sound}, developing tools for steering music AI with various constraints~\cite{louie2020novice}, and integrating musical AIs into co-design processes for therapeutic applications~\cite{sun2024understanding}. The interaction involves manipulating acoustic properties and structures, enabling novel forms of sonic expression and exploration.

\textbf{\textcolor{color1}{Modality-4. Multimodal and Hybrid Interaction:}} Many co-creative endeavors integrate multiple modalities or involve complex artifact creation that transcends a single interaction channel. This category encompasses the collaborative making of physical or digital products~\cite{zhang2023towards,vear2024jess+,benjamin2023entoptic,liu2024he,brand2023envisioning,kang2024design}, interactive installations~\cite{cremaschi2024steampunk,trajkova2024exploring}, and the co-design of user interfaces~\cite{ayobi2023computational}, experiences~\cite{claisse2024understanding,zheng2024charting,chang2024co}, systems~\cite{kamath2024sound,lin2021engaging,gmeiner2023exploring}, and clinical decision support tools~\cite{zhang2022you}. AI might contribute through design generation, simulation, user data analysis~\cite{trajkova2024exploring,wan2024metamorpheus}, prototyping~\cite{kamath2024sound,bauer2024musictraces}, or providing feedback~\cite{shaer2024ai}, complementing human insight. Research in this area explores critical perspectives on AI's role~\cite{cremaschi2024steampunk,benjamin2023entoptic}, develops design principles for generative AI interfaces~\cite{weisz2024design}, investigates tools for specific contexts like triage autonomy~\cite{berge2023designing} or community health~\cite{claisse2024understanding}, enhances prototyping methods~\cite{wang2024farsight}, designs multi-sensory collaborative systems~\cite{bauer2024musictraces}, explores embodied AI collaboration~\cite{zhang2023towards,vear2024jess+}, and leverages techniques like LLM-enhanced brainwriting~\cite{shaer2024ai}. This modality often involves intricate interplay between different forms of input and output, supporting complex design and making processes~\cite{zhang2022get,kang2024design}.

\begin{figure}[!htbp]
    \includegraphics[width=1\textwidth]{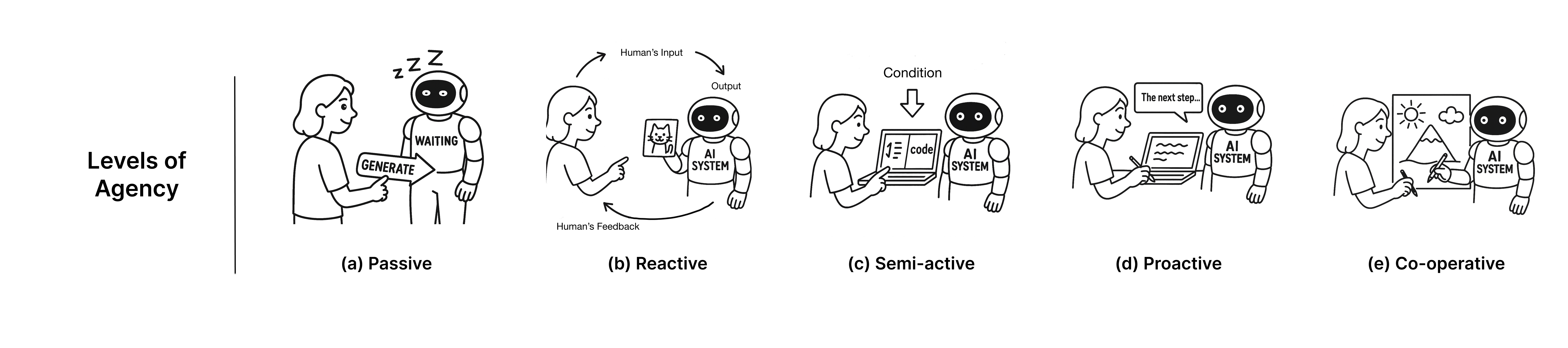}
    \caption{The illustration of different levels of agency: (a)~\cite{suh2021ai}, (b)~\cite{newman2024want}, (c)~\cite{weisz2022better}, (d)~\cite{hwang2022too}, (e)~\cite{ayobi2023computational}. The illustrations were generated by ChatGPT and modified by authors.}
\end{figure}

\section{Agency: Patterns and Distribution}

\textbf{Of the 134 papers identified, for the classification of agency, we specifically focused on the 106 papers that presented concrete system implementations or detailed interaction designs. This selection criterion was crucial because assessing agency types requires a clear description of how humans and AI interact within a defined system.} The remaining 28 papers were excluded from this specific analysis as they primarily discussed theoretical concepts, users studies without sufficiently detailing the underlying system, or lacked the necessary implementation details to allow for a classification of agency dynamics.

\subsection{Agency Patterns}\label{sec:agency_patterns}
\begin{figure}[h]
    \centering
    \includegraphics[width=0.4\textwidth]{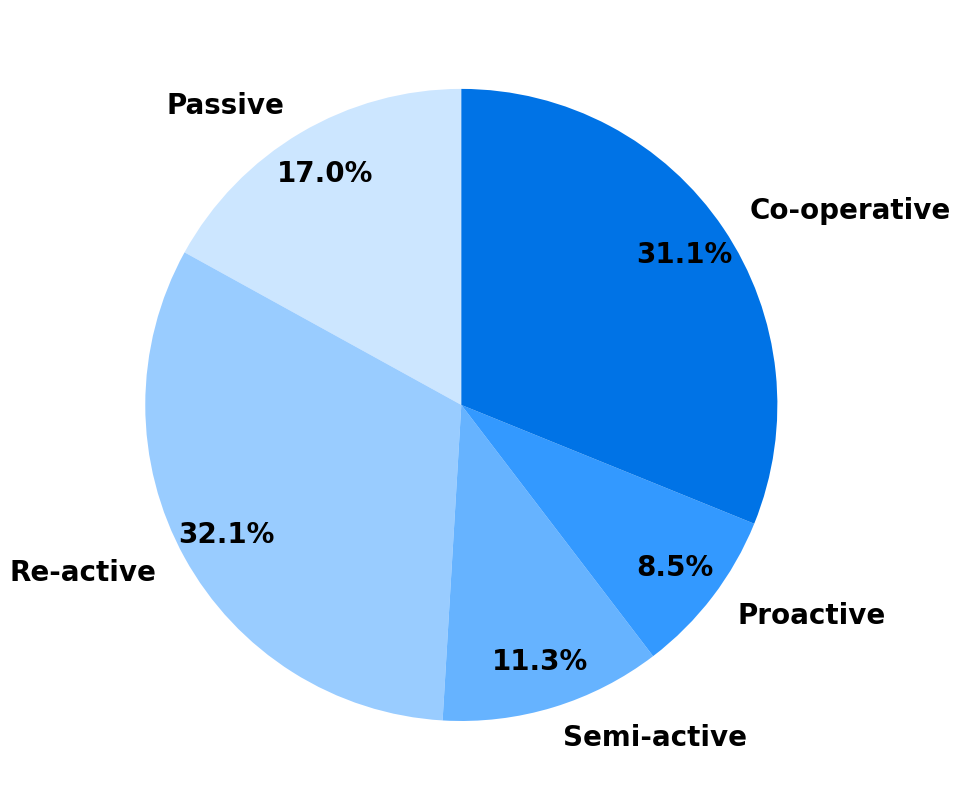} 
    \caption{Distribution of AI Agency Patterns.}
    \label{fig:distribution} 
\end{figure}

To understand how AI agency manifests in co-creative systems, we analyzed the 106 selected papers based on the AI's role and initiative during interaction, drawing conceptually from frameworks classifying AI behavioral patterns \cite{ rammert2008action}. This approach focuses on the nature of the AI's contribution to the interaction flow, ranging from passive assistance to proactive partnership. Our analysis revealed a spectrum of AI agency patterns across the reviewed systems. The distribution (Figure~\ref{fig:distribution}, using the terminology from \cite{rammert2008action}), was as follows: 18 systems exhibited \textbf{Passive} agency (acting only upon direct user invocation), 34 demonstrated \textbf{Reactive} agency (responding directly to user input/actions), 12 featured \textbf{Semi-active} agency (initiating actions under specific conditions), 9 showed \textbf{Proactive} agency (taking independent initiative), and 33 involved \textbf{Co-operative} agency (acting collaboratively alongside the user). These patterns are detailed below with specific examples:

\textbf{\textcolor{color2}{Pattern-1. Co-operative Agency:}} Research demonstrates co-operative agency through AI collaborating with humans to enhance various processes and outcomes. Examples of co-operative agency include AI tools identifying potential harms during prototyping~\cite{wang2024farsight}, systems aiding academic meta-review by summarizing peer reviews~\cite{sun2024metawriter}, algorithms enabling users to aesthetically personalize prosthetics~\cite{zhou2023beyond}, and computational notebooks using interactive visualizations for shared stakeholder understanding of healthcare ML models~\cite{ayobi2023computational}.

\textbf{\textcolor{color2}{Pattern-2. Proactive Agency:}} Proactive agency is when AI takes initiative or introduces unique insights early in co-creation. Examples include: AI offering unique ``non-human'' perspectives that users might not reach alone, even if this potential isn't fully tapped yet~\cite{hwang2022too}, AI proactively providing next-sentence or next-paragraph suggestions to steer the writing process~\cite{dhillon2024shaping}, generative AI tools initiating the creation of novel sound options for designers~\cite{kamath2024sound}, and interactive tools proactively highlighting relevant news or generating specific use cases to make users consider potential AI harms early on~\cite{wang2024farsight}.

\textbf{\textcolor{color2}{Pattern-3. Semi-active Agency:}} Semi-active agency involves AI providing support when requested by users. Examples include AI performing code translation upon a user's request~\cite{weisz2022better}. Researchers also designed AR applications where features likely activate based on user interaction to enhance social experiences~\cite{reig2023supporting}. Others demonstrated computational plug-ins for design software, allowing users to actively call upon machine learning tools when needed~\cite{hegemann2023computational}. Additionally, peer-mentoring systems were developed, presumably offering students support or connections upon their engagement with the system~\cite{anthraper2024peerconnect}.

\textbf{\textcolor{color2}{Pattern-4. Re-active Agency:}} Re-active agency is when AI responds directly to user actions or behaviors. For example, Gmeiner et al. developed AI design tools react to designer inputs, highlighting interaction challenges~\cite{gmeiner2023exploring}. Lukoff et al. designed tools with different reactions, including implicit responses towards autoplay and explicit ones towards searching~\cite{lukoff2021design}. Personalization systems, such as news recommenders, also operate reactively based on user data~\cite{rezk2024agency}. This reactive function pervasive also for generative AI, which responds directly to children's prompts~\cite{newman2024want}.

\textbf{\textcolor{color2}{Pattern-5. Passive Agency:}} Passive agency describes AI's subtle influence on human dynamics. Examples include: AI implicitly shaping musicians' collaborative dynamics by affecting common ground and roles~\cite{suh2021ai}, generative AI impacting educational interactions through generated content while passively raising concerns about authorship and bias~\cite{han2024teachers}, and AI being used within therapist-guided family storymaking sessions, setting a context where it might subtly influence group interactions~\cite{liu2024he}.

\subsection{Agency Distribution}\label{sec:agency_distribution}

A foundational aspect of human-AI co-creation is the \textbf{distribution of agency}, defined as the distribution and dynamics of decision-making authority between human and AI agents. Extending the work from Holter et al.~\cite{holter2024deconstructing}, which classified agency along distribution (human, mixed, AI) and allocation (pre-determined, negotiated) dimensions, we incorporate these and add granularity as a third key aspect. Effectively designing and analyzing such systems requires a framework that comprehensively captures how agency is structured. We propose characterizing agency distribution through three fundamental, orthogonal dimensions: the \textbf{locus} (determining who holds primary decision-making authority), the \textbf{dynamics} (addressing how this authority is managed or shifted over time and interaction), and the \textbf{granularity} (specifying the level of abstraction or detail at which authority is exercised). 

\textbf{\textcolor{color2}{Dimension-1. Locus:}} This dimension specifies where primary decision-making authority resides along a spectrum. In human-centric control, humans retain ultimate authority, with AI serving assistive or advisory roles \cite{lee2024and}. Conversely, AI-centric control involves autonomous AI execution based on defined objectives, common in full automation \cite{tholander2023design}. Between these poles, hybrid or shared control models distribute authority (e.g., as collaborators \cite{louie2020novice, anthraper2024peerconnect}), significantly impacting interaction dynamics and user perception. Precisely defining this locus gains complexity and relevance as AI capabilities evolve from tools to more autonomous agents.

\textbf{\textcolor{color2}{Dimension-2. Dynamics:}} This dimension involves two primary modes for managing control authority during interaction. \textbf{Static allocation} assigns agency based on fixed rules, roles, or predefined stages determined during system design, offering predictability. Examples include specifying distinct human-AI cooperative roles \cite{lin2024text}, implementing predetermined control levels through interface design \cite{sharma2024generative}, or structuring interaction based on distinct task phases \cite{sun2024understanding}. \textbf{Dynamic allocation} determines control situationally during the interaction itself, adapting to evolving contexts but posing significant challenges in negotiation and alignment \cite{dudley2018review}. Such dynamics are evident when interaction control shifts based on evolving factors, like the quality of AI contributions influencing a collaborative process \cite{weisz2022better}.

\textbf{\textcolor{color2}{Dimension-3. Granularity:}} This dimension defines the level of detail at which agency is exercised. Control can operate at a \textbf{high level}, concerning strategic goals or overall workflow \cite{gu2024data}, or at a \textbf{fine-grained level}, involving specific actions or parameter adjustments within a process \cite{zhou2024understanding}. Selecting appropriate granularity is crucial for complex tasks \cite{suh2024luminate}, as finer control points can enhance user agency and transparency, mitigating ``black box'' issues by enabling steerable interactions \cite{moruzzi2024user}. Examples include implementing varying levels of AI scaffolding, like sentence versus paragraph suggestions \cite{dhillon2024shaping}, or providing distinct control types, such as domain-specific versus technology-specific parameters in generation \cite{kamath2024sound}.

\section{Control Mechanisms}\label{sec:control_mechanisms}

We first broadly categorized four types of processes including control according to the Input-Process-Output-Process (IPOF) model~\cite{wiener2019cybernetics}, forming a space ranging from human-initiated methods to AI-initiated methods~\cite{saint2021machine}. We then summarized three types of control mechanism within each process. We opted not to use other co-creation processes because not all tasks and mechanisms emerge in the stages for these classifications~\cite{kabirunleashing}. For example, Kabir et al.~\cite{kabirunleashing} classified the co-creation process to contain preparation, exploration, collaboration, development, implementation and evaluation. Wu et al.~\cite{wu2021ai} classified the human-AI co-creation as perceive, think, express, collaborate, build and test.  

\subsection{Input: Information Access}

\textbf{\textcolor{color3}{Mechanism-1. Guided Input Interaction:}} This mechanism encompasses methods that structure and facilitate how users provide input to AI systems. It includes user-centered optimization, interface-supported guidance and multimodal input integration. \textbf{User-centered input optimization} enhances users control to align AI outputs with expectations. Users actively refine prompts iteratively based on AI responses~\cite{lin2024text,tholander2023design} or tailor their experience via customizable interface elements like adjustable parameters and views~\cite{moruzzi2024user,hoque2024hallmark}. Users can also strategically design prompts to shape AI behavior or manipulate bias for specific goals~\cite{sharma2024generative,cremaschi2024steampunk}.  \textbf{Interface-supported input guidance} employs interface design to actively direct user input, improving interaction efficiency. Interfaces provide interactive visual feedback (e.g., confidence highlighting, code comparison) to help users understand and refine inputs~\cite{weisz2022better,albaugh2023augmented}. Techniques like multi-panel layouts with progressive disclosure organize complex information~\cite{liu2024coquest,wang2024farsight}, while context-aware conversational interfaces reduce cognitive load with intuitive response mechanisms~\cite{fan2024contextcam,fu2024text}.   \textbf{Multimodal input integration} incorporates diverse input types to enrich interaction. Systems map user gestures or movements to algorithmic parameters~\cite{zhou2023beyond,gebreegziabher2023patat}, create synergy between specialized tools and interfaces for cohesive input environments~\cite{verheijden2023collaborative,ayobi2023computational}, or utilize cross-platform approaches to facilitate collective AI guidance~\cite{anthraper2024peerconnect,dhillon2024shaping}.

\textbf{\textcolor{color3}{Mechanism-2. Context Awareness and Memory Retention:}} This mechanism enables AI systems to leverage historical, environmental, task, and user information, enhancing contextual understanding and collaboration continuity, encompassing several detailed approaches. For \textbf{interaction history integration}, systems incorporate past conversational or operational history into decision-making~\cite{sharma2024generative,fu2024text}. Process tracking features preserve provenance, helping users follow idea evolution or reflect on progress~\cite{liu2024coquest,anthraper2024peerconnect}. For \textbf{environmental and spatial awareness}, AI recognizes and adapts to physical or digital surroundings. Systems may track user position and movement to adjust interactions~\cite{praveena2023exploring,bauer2024musictraces} or collect real-time contextual data (e.g., location, weather, sensor data) to provide relevant recommendations~\cite{fan2024contextcam}. For \textbf{task-oriented context management}, systems maintain awareness of specific tasks, objectives, and histories to ensure consistency. They might enable retrieving artifacts for context-aware iteration~\cite{verheijden2023collaborative}, track design goals~\cite{gmeiner2023exploring}, adapt interfaces to workflow stages~\cite{zhang2022get}, or use history to inform alerts~\cite{wang2024farsight}. For \textbf{personalized context adaptation}, AI tailors its behavior to individual user backgrounds, needs and preferences. This includes adapting content for specific users~\cite{ayobi2023computational}, designing culturally responsive activities~\cite{li2023want}, considering community sensitivities~\cite{bircanin2021including} or reflecting situated literacies~\cite{benjamin2023entoptic}.

\textbf{\textcolor{color3}{Mechanism-3. Transparency and Explainability:}} this mechanism enhance user understanding of AI operations and build trust. They range from tracking concrete operations to prompting reflection on abstract concepts, encompassing four main approaches. For \textbf{interaction tracking}, systems document and present interaction histories, allowing users to review contributions from both humans and AI. This clear record supports trust. For example, Hoque et al.~\cite{hoque2024hallmark} implement history tracking for writers to manage AI contributions, while Anthraper et al.~\cite{anthraper2024peerconnect} visualize progress bars and goal achievement in a mentoring system. For \textbf{decision visualization}, interfaces visualize AI's decision-making processes or influencing factors understandably. For instance, Berge et al.~\cite{berge2023designing} provides nurses with clear interface suggestions highlighting urgent symptom combinations, and Zhou et al.~\cite{zhou2023beyond} make algorithmic processes transparent by visually linking dancers' actions to evolving design results. For \textbf{system explanation}, these systems explain the reasoning behind AI-generated outputs to clarify internal workings. Examples include providing rationales for generated research questions~\cite{liu2024coquest}, documenting iterative project changes to track solution evolution~\cite{zheng2024charting}, surfacing relevant incident reports to explain potential harms~\cite{wang2024farsight}, or explaining AIMC tool suggestions~\cite{fu2024text}. For \textbf{ideological reflection}, designs prompt users to reflect critically on technology's underlying ideologies and opacity. Cremaschi et al.~\cite{cremaschi2024steampunk}, for example, juxtapose a typewriter with modern AI to provoke thoughts on transparency, while Benjamin et al.~\cite{benjamin2023entoptic} use a camera-based `entoptic metaphor' enabling non-technical users to explore AI's influence.

\subsection{Action: Exploration}

\textbf{\textcolor{color3}{Mechanism-4. Multimodal Action Space Exploration:}} This mechanism facilitates human-AI collaboration by enabling interaction through diverse modalities beyond traditional text and graphics. Systems explore actions using various channels: \textbf{Text-based interaction} remains fundamental, supporting communication tools~\cite{fu2024text} or integrating text feedback with graphical interfaces~\cite{liu2024coquest}, though differences from visual interaction are noted~\cite{tholander2023design}. \textbf{Visual interaction} employs graphical elements, such as visual body representations for symptom localization~\cite{berge2023designing} or interfaces combining text prompts with sketch editing for image generation~\cite{verheijden2023collaborative,shi2020emog}. \textbf{Multi-sensory and physical interaction} integrates rich channels. Systems may use visual, auditory and kinesthetic elements~\cite{ayobi2023computational}: visual supports with tactile experiences~\cite{bircanin2021including}, map physical movements via motion capture~\cite{zhou2023beyond}, or incorporate full-body actions, gestures, smart objects, and haptics~\cite{albaugh2023augmented,bauer2024musictraces,capel2023human}. \textbf{Combined and integrated modalities} explicitly merge interaction modes. Systems might integrate goal-setting, messaging, and calendars~\cite{anthraper2024peerconnect}: combine sound generation with visual feedback~\cite{kamath2024sound}, support diverse inputs like sketches and gestures~\cite{weisz2024design}, or integrate various data types (e.g., text, voice, photos, sensor data) for comprehensive context~\cite{fan2024contextcam,wang2024farsight}.

Within these applications, systems regulate agency through strategies like combining modalities for user choice~\cite{verheijden2023collaborative}, integrating multiple sensory channels~\cite{ayobi2023computational}, mapping physical actions directly to controls~\cite{zhou2023beyond}, and adapting interaction methods based on context~\cite{fan2024contextcam}. 
\textbf{\textcolor{color3}{Mechanism-5. Action Coordination:}} This mechanism addresses how human-AI co-creation systems distribute responsibilities, decision-making authority, and interaction dynamics based on the assumed roles of human and AI agents. Five primary patterns emerge: For \textbf{complementary role distribution,} systems assign distinct, interdependent roles leveraging respective strength. Humans typically provide strategic direction or creative input, while AI handles data processing or routine tasks~\cite{hoque2024hallmark,liu2024coquest,gmeiner2023exploring,anthraper2024peerconnect,zhang2022get}. For \textbf{human-dominated agency with AI support,} humans retain primary decision-making authority, utilizing AI as an auxiliary tool or assistant that provides suggestions or analysis but lacks autonomous agency~\cite{verheijden2023collaborative,bircanin2021including,li2023want,dhillon2024shaping,fu2024text}. For \textbf{shared creative agency,} humans and AI engage jointly with mutual influence on the creative output. Both participate throughout idea generation, refinement and evaluation, fostering an emergent process~\cite{zheng2024charting,zhou2023beyond,ayobi2023computational,capel2023human,cremaschi2024steampunk}. For \textbf{technical precision and control,} systems focus on enabling users to exert detailed, fine-grained control over AI outputs through interfaces offering parameter tuning, prompt refinement, or direct manipulation~\cite{kamath2024sound,shi2020emog,wang2024farsight,gebreegziabher2023patat,fan2024contextcam}. For \textbf{autonomous AI contribution with human frameworks}, AI operates with significant autonomy on specific sub-tasks within parameters established by humans, contributing independently while remaining accountable to human oversight~\cite{shaer2024ai,praveena2023exploring,bauer2024musictraces,weisz2024design,hwang2022too}.

\textbf{\textcolor{color3}{Mechanism-6. Attention-focused Processing:}}
Attention mechanisms direct the system's focus to specific parts of the data or the processing pipeline. For example, in a machine-translation system, attention mechanisms can focus on certain words or phrases in the source text to generate a more accurate translation output. By allocating more processing resources to these key elements, the system can improve the quality of its actions and outputs. This mechanism is particularly useful in complex tasks where certain aspects of the input data are more critical than others. Hoque et al.~\cite{hoque2024hallmark} uses interface design to focus writer attention on key areas like text editing and interaction history. Systems like PromptCharm~\cite{wang2024promptcharm} use cross-attention techniques and interactive interfaces to visualize attention distributions, further enhancing explainability and user engagement.

\subsection{Output: Direct Intervention}

\textbf{\textcolor{color3}{Mechanism-7. Modification and Intervention:}} This mechanism detail how users exert control over AI systems by intervening in processes or modifying outputs. We observed four main approaches. For \textbf{direct editing and adjustment}, user directly alter AI outputs. This includes manually labeling or modifying generated text~\cite{hoque2024hallmark}, identifying and correcting errors in AI-generated code~\cite{weisz2022better}, or manipulating visual design elements through gestures~\cite{zhou2023beyond}. For \textbf{parameter and prompt control}, users influence AI outcomes indirectly through system settings or inputs, rather than altering the output itself. Examples include shaping generative processes via prompts and parameters~\cite{weisz2024design}, controlling the sequence or level of AI assistance~\cite{dhillon2024shaping}, or adjusting AI decision-making parameters~\cite{capel2023human}. For \textbf{real-time intervention and adjustment,} users intervene during an ongoing AI process. Cinematographers might adjust robot paths live~\cite{praveena2023exploring}, developers may integrate real-time alerts into workflows~\cite{wang2024farsight}, or artists can intervene mid-creation by adding strokes or selecting suggestions~\cite{shi2020emog}. For \textbf{acceptance or rejection or AI suggestions,} users act as gatekeepers by explicitly accepting or rejecting AI contributions. This includes users overriding suggestions and documenting their own assessments~\cite{berge2023designing}, clinicians manually acknowledging or dismissing system-identified problems~\cite{zhang2022get}, or users controlling code annotations by accepting/rejecting recommendations~\cite{gebreegziabher2023patat}.
\textbf{\textcolor{color3}{Mechanism-8. Adaptive Scaffolding:}} This mechanism dynamically adjusts the level and type of AI assistance provided to users, operating on a spectrum from system-controlled to user-driven approaches, often incorporating hybrid models. For \textbf{system-controlled adaptive scaffolding,} the AI autonomously modifies support based on pre-defined rules, learned models, or analysis of user behavior and context. Examples include AI suggesting relevant questions based on conversation flow~\cite{berge2023designing}, adjusting guidance based on analyzed user progress~\cite{verheijden2023collaborative,weisz2024design}, adapting assistance levels to designer skill~\cite{gmeiner2023exploring}, tailoring support based on learner understanding~\cite{li2023want}, structuring learning based on observed behavior~\cite{ayobi2023computational}, adapting based on user interaction interpretation~\cite{liu2024coquest}, or curating data views based on clinical context~\cite{zhang2022get}. For \textbf{user-driven adaptive scaffolding,} users explicitly control the adaptation of support mechanisms. They might manually adjust settings, select different assistance levels~\cite{albaugh2023augmented}, request specific types of support~\cite{anthraper2024peerconnect}, trigger or dismiss system hints~\cite{bauer2024musictraces}, specify proficiency levels~\cite{dhillon2024shaping}, or set goals to modify assistance intensity~\cite{fu2024text}. For \textbf{hybrid adaptive scaffolding,} these approaches combine system autonomy with user control, where the system might make adjustments but allow user overrides or modifications. Adaptation factors include user proficiency~\cite{berge2023designing,dhillon2024shaping}, task phase or context~\cite{verheijden2023collaborative,fu2024text}, explicit user feedback~\cite{liu2024coquest,albaugh2023augmented}, and observed task progress~\cite{ayobi2023computational,weisz2024design}.

\textbf{\textcolor{color3}{Mechanism-9. Chain-of-Thought:}} This mechanism directly intervenes in the output phase to explicitly display the AI's step-by-step reasoning process used to reach a solution or suggestion. Making the system's logic transparent helps users understand and evaluate the output. For instance, Liu et al.~\cite{liu2024coquest} employ CoT prompting to improve LLM reasoning and make the thought process visible when generating research questions. Zheng et al.~\cite{zheng2024charting} use CoT documentation to analyze reasoning behind AI suggestions, fostering deep evaluation. Others mention CoT as a technique for providing rationales for AI outputs~\cite{weisz2024design}, or use few-shot CoT approaches to enhance LLM problem-solving within specific contexts like ContextCam~\cite{fan2024contextcam}.

\subsection{Feedback: Stressing}

\textbf{\textcolor{color3}{Mechanism-10. Confidence Visualization:}} This mechanism communicate AI reliability or influence user confidence regarding AI outputs and actions. Systems employ several approaches: For \textbf{confidence/uncertainty visualization,} interface visually represent the AI's confidence level (e.g., via scores, intervals or alternatives) for its outputs or actions~\cite{weisz2024design}. This helps users gauge reliability and make informed decisions about interpreting or acting upon AI contributions. For \textbf{user feedback and trust management,} systems incorporate feedback channels to measure how design choices influence users' perception of system reliability, bias, and overall trust~\cite{sharma2024generative}. For \textbf{confidence-based ranking and prioritization,} algorithms order or rank AI suggestions based on calculated confidence metrics. This guides user attention towards reliable items first, as seen in pattern prioritization based on model confidence~\cite{gebreegziabher2023patat}. For \textbf{user self-efficacy enhancement,} some systems include psychological or social support mechanisms designed to boost users' confidence. For example, PeerConnect aims to increase students' self-efficacy and sense of belonging, thereby enhancing confidence in their engagement~\cite{anthraper2024peerconnect}.

\textbf{\textcolor{color3}{Mechanism-11. Explanatory Feedback Emphasis:}} this mechanism employs transparency strategies to help users understand AI operations and reasoning, facilitating informed decisions via targeted approaches. For \textbf{model explanation and reasoning disclosure}, systems provide explicit insights into AI decision-making. This involves interpreting ambiguous outputs~\cite{lin2024text,tholander2023design}, offering views into model operations~\cite{moruzzi2024user}, using visualizations like feature importance plots~\cite{ayobi2023computational}, explaining design outcomes~\cite{gmeiner2023exploring}, displaying flags based on clinical rules~\cite{zhang2022get}, or providing interpretive metaphors~\cite{benjamin2023entoptic}. For \textbf{visual highlighting and differentiation,} interfaces use visual cues to distinguish AI contributions or guide user attention. Examples include color-coding AI-written text~\cite{hoque2024hallmark}, highlighting key points in peer reviews~\cite{sun2024metawriter}, emphasizing generative variability~\cite{weisz2024design}, or visually marking text portions matching code patterns~\cite{gebreegziabher2023patat}. For \textbf{user control and interactive transparency,} interfaces allow users to interactively explore AI behavior or control processes. Learners might see the direct impact of their decisions~\cite{li2023want}, users might examine training data and methods~\cite{hwang2022too}, track machine states during fabrication~\cite{albaugh2023augmented}, or review incident reports explaining potential harms~\cite{wang2024farsight}. For \textbf{contrast and metaphorical representation,} designs may use contrast or metaphors to explain AI operations or highlight opacity. This includes designs aiming for process clarity~\cite{verheijden2023collaborative}, systems providing understandable results based on clear inputs~\cite{shi2020emog}, or contrasting old and new technologies to critique opacity~\cite{cremaschi2024steampunk}.
\textbf{\textcolor{color3}{Mechanism-12. Iterative Feedback Loop:}} This mechanism enables continuous refinement of AI systems and co-created outputs through dynamic exchanges based on user input or environmental changes. Feedback can be primarily user-directed, system-initiated or bidirectional. For \textbf{user-directed feedback}, users explicitly provide feedback to guide AI performance and improvements. Examples include using AI outputs to inform subsequent prompts~\cite{lin2024text}, offering direct interaction feedback or ratings~\cite{moruzzi2024user,hoque2024hallmark,berge2023designing}, providing textual feedback on generated content~\cite{liu2024coquest}, adjusting movements based on real-time visualizations~\cite{zhou2023beyond}, iteratively testing outputs with parameter adjustments~\cite{gmeiner2023exploring}, manually indicating alignment with assessments~\cite{zhang2022get}, refining computational methods~\cite{hegemann2023computational}, or correcting model outputs for retraining~\cite{gebreegziabher2023patat}. For \textbf{system-initiated feedback}, these systems proactively provide these feedback to enhance the interaction process or guide users. AI might respond to users' additional queries~\cite{tholander2023design}, provide real-time suggestions with system adaptations~\cite{shi2020emog}, use generated sounds for rapid iteration~\cite{kamath2024sound}, offer immediate visual feedback via interfaces~\cite{praveena2023exploring}, monitor user actions and adjust responses~\cite{bauer2024musictraces,albaugh2023augmented}, serve as a feedback tool during ideation~\cite{shaer2024ai}, track goals for immediate feedback~\cite{anthraper2024peerconnect}, propose updated content based on input~\cite{fan2024contextcam}, or establish user-system response cycles for refinement~\cite{wang2024farsight,cremaschi2024steampunk,verheijden2023collaborative,zheng2024charting,sharma2024generative}. For \textbf{bidirectional interaction feedback}, the feedback flows dynamically in both directions between users and the AI. This occurs through iterative refinement in workshops~\cite{ayobi2023computational}, designers interpreting non-verbal participant behaviors~\cite{bircanin2021including}, iterative design sessions involving discussion~\cite{li2023want}, continuous feedback mechanisms~\cite{capel2023human}, or incorporating user experiences into development~\cite{benjamin2023entoptic}.

\section{Applications}

While Wu et al.~\cite{wu2021ai} categorize AI creativity applications broadly (e.g., Culture, Industry), our review's specific focus on HCI literature and co-creative agency patterns reveals granular domains emerging from the analyzed papers. We therefore detail how agency and control mechanisms are operationalized within these application domains, analyzing the commonalities and limitations of agency frameworks.  

\textbf{\textcolor{color4}{Application-1. News:}} Agency dynamics in news curation typically involve \textit{semi-proactive} and \textit{reactive} mechanisms. Dhillon et al.~\cite{dhillon2024shaping} reveal tensions between AI-augmented cognitive support (automatic) and authorial control (reactive), where writers actively revise AI suggestions. Hoque et al.~\cite{hoque2024hallmark} employ reactive provenance tracking via HALLMARK to ensure transparency in LLM interactions. Rezk et al.~\cite{rezk2024agency} identify a \textit{passive} behavioral-intention gap in news recommenders, where users desire agency but avoid active intervention. Sharma et al.~\cite{sharma2024generative} test \textit{proactive} exposure through biased LLMs that either reinforce or challenge user viewpoints. However, when AI offers automated cognitive support, yet writers retain final authorial control through active revision, the tension is evident~\cite{dhillon2024shaping}.

\textbf{\textcolor{color4}{Application-2. Healthcare:}} Medical AI employs \textit{expert-supervised reactive} agency, where clinicians maintain final decision-making authority (\textit{reactive}). Berge et al.~\cite{berge2023designing} emphasize \textit{semi-proactive} workflow integration to balance AI decision-support with documentation demands. Claisse et al.~\cite{claisse2024understanding} adopt \textit{participatory proactive} design for culturally sensitive mHealth tools. Sun et al.~\cite{sun2024understanding} implement \textit{modality-specific reactive control} in music therapy, while Zhou et al.~\cite{zhou2023beyond} use \textit{creative semi-proactive} co-production for prosthetic design.

\textbf{\textcolor{color4}{Application-3. Art:}} \textit{Tool-mediated passive} and \textit{semi-proactive} agency dominate here. Lin et al.~\cite{lin2024text} show how \textit{prompt engineering} (\textit{reactive}) enhances design ideation through forces verbal articulation. Tholander et al.~\cite{tholander2023design} highlight \textit{interface-driven reactive} control affecting user expectations. Kamath et al.~\cite{kamath2024sound} deploy \textit{mixed-initiative semi-proactive} systems with iterative refinement capabilities. 

\textbf{\textcolor{color4}{Application-4. Education \& Research:}} Educational AI systems integrate \textit{curriculum-bound reactive} and \textit{proactive} mechanisms. Weisz et al.~\cite{weisz2022better} demonstrate \textit{error-correction reactive} agency in code tools, while Gebreegziabher et al.~\cite{gebreegziabher2023patat} design pattern-interpretable proactive systems like PaTAT. Liu et al.~\cite{liu2024coquest} contrast \textit{exploration strategy proactive} (breadth-first) and \textit{reactive} (depth-first) approaches in research question generation. Sun et al.~\cite{sun2024metawriter} use \textit{context-aware semi-proactive} meta-review systems. 

\textbf{\textcolor{color4}{Application-5. Entertainment:}} Creative industries employ \textit{vision-aligned proactive} and \textit{semi-proactive} agency. Fan et al.~\cite{fan2024contextcam} develop \textit{context-aware proactive} co-creation systems integrating environmental data. Louie et al.~\cite{louie2020novice} implement \textit{slider-based reactive} controls for music parameters. Verheijden et al.~\cite{verheijden2023collaborative} demonstrate \textit{collaborative semi-proactive} editing via BrainFax's chatbot interfaces. 

\textbf{\textcolor{color4}{Application-6. Software Development:}} Development workflows prioritize \textit{value-aligned proactive} strategies. Varanasi et al.~\cite{varanasi2023currently} identify \textit{proactive value lever} strategies for ethical alignment. Wang et al.~\cite{wang2024farsight} introduce \textit{harm-aware reactive} control through FARSIGHT's incident-linked prototyping tools. 

\textbf{\textcolor{color4}{Application-7. Accessibility:}} Assistive technologies emphasize \textit{empowerment proactive} and \textit{semi-proactive} mechanisms. Shen et al.~\cite{shen2022kwickchat} design \textit{keyword-driven proactive} communication systems for motor-impaired users. Bircanin et al.~\cite{bircanin2021including} operationalize \textit{participatory proactive} principles like ``maximizing choice'' to inform inclusive HCI methods.  

\begin{figure}[!htbp]
    \includegraphics[width=1\textwidth]{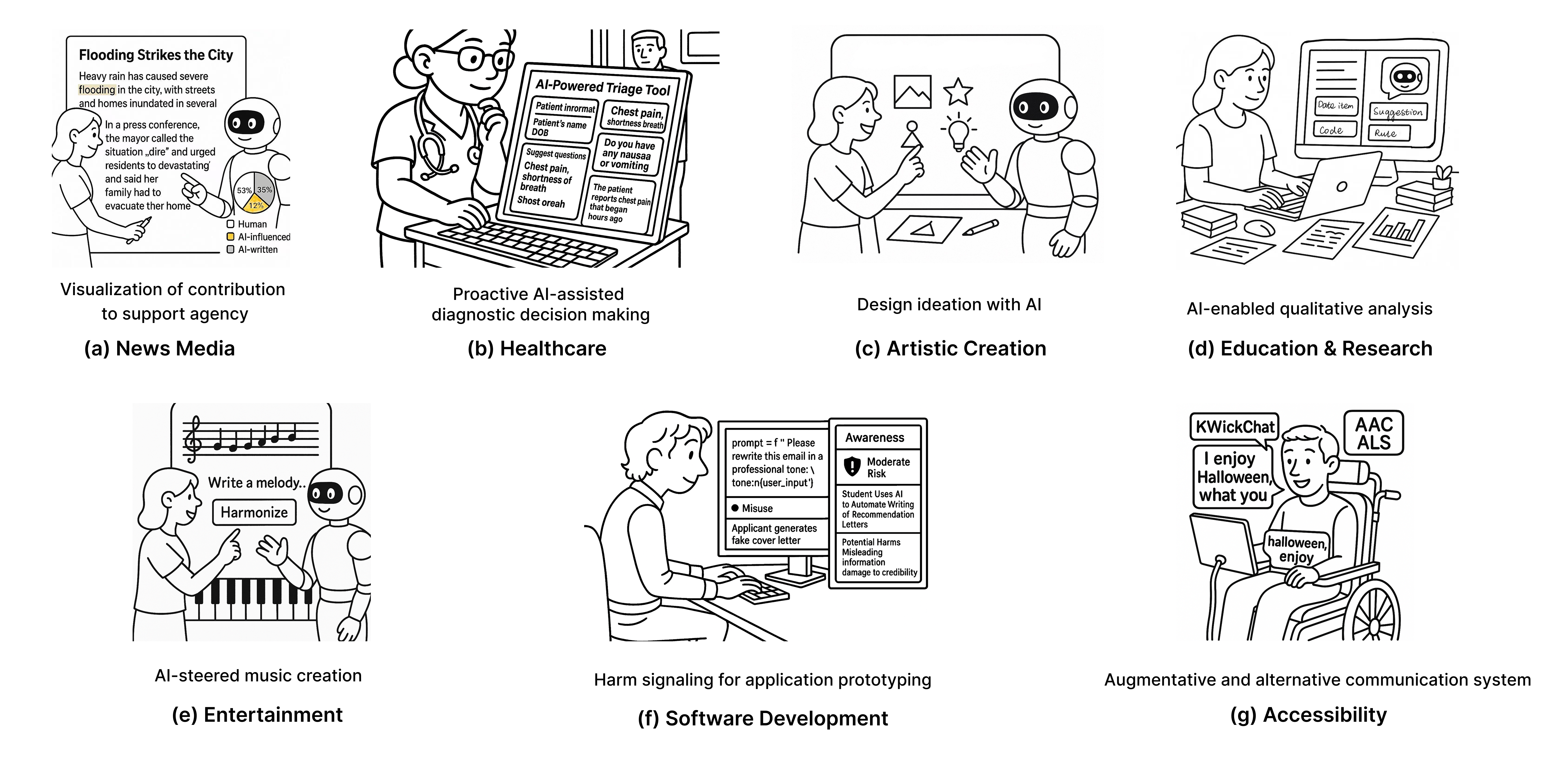}
    \caption{Illustration of human-AI co-creation applications across different domains: (a)~\cite{hoque2024hallmark}, (b)~\cite{berge2023designing}, (c)~\cite{tholander2023design}, (d)~\cite{gebreegziabher2023patat}, (e)~\cite{louie2020novice}, (f)~\cite{wang2024farsight}, (g)~\cite{shen2022kwickchat}. The illustrations were generated by ChatGPT and modified by authors.}
\end{figure}

\section{Challenges \& Directions}

We structure these co-creation challenges into: \textbf{collaborative experience} (Challenges 1, 2) concerning creativity, ownership, and trust, \textbf{collaboration infrastructure} (Challenges 3, 4) addressing data security and system interoperability, and \textbf{social implications} (Challenges 5, 6, 7) covering equity, ethics, and future paradigms. These research directions guide the design of responsible human-AI collaborative systems.

\textbf{\textcolor{color5}{Challenge-1. Creativity and Ownership:}} While enhancing human creativity with suggestions and process facilitation, AI simultaneously complicates ownership and intellectual property rights. Distinguishing human input from AI contributions grows critical yet complex. Research indicates co-creation raises unique authorship questions and copyright concerns, distinct from solely AI outputs~\cite{sun2024generative}. Kamath et al.~\cite{kadoma2024role} argue that AI completing tasks, instead of humans, potentially blurs authorship and diminishes users' sense of agency. Consequently, developing clear legal frameworks and guidelines for co-created content becomes imperative. System designers must prioritize user control and perceived ownership, designing interactions that foster collaboration while preserving human agency. . Although current studies investigate factors influencing ownership feelings~\cite{guo2024exploring}, fully resolving the intertwined legal, ethical, and experiential dimensions of ownership in human-AI co-creation demands significant ongoing effort and innovation.

\textbf{\textcolor{color5}{Challenge-2. Trust and Transparency:}} Building user trust necessitates AI systems with transparent decision-making processes and clear mechanisms for user influence or correction. Yet, significant obstacles impede this goal; AI's role within teams is often ambiguous, and its "inhuman" reasoning patterns challenge user comprehension and acceptance~\cite{hwang2022too}. This lack of clarity can correlate with broader distrust, including data privacy concerns~\cite{rezk2024agency}. Although approaches like explainable AI offer promise in demystifying AI actions, rendering complex models truly understandable remains a persistent difficulty, potentially undermining user confidence. Consequently, balancing the operational autonomy of AI with sufficient human oversight is crucial for cultivating and sustaining user confidence, representing a core design challenge.

\textbf{\textcolor{color5}{Challenge-3. Data Privacy and Security:}} AI's reliance on extensive data creates a fundamental tension with the imperative to protect user privacy and security, especially when processing sensitive information like children's~\cite{zhang2024mathemyths}. Effectively safeguarding user data demands robust encryption and access controls, coupled with mechanisms granting users clear control and transparency regarding data usage~\cite{wang2023treat}, including opt-in/out choices to foster trust. Despite these safeguards, striking the necessary balance between leveraging vast datasets for optimal personalized AI performance (and for monetization~\cite{wang2023treat}) and upholding privacy standards constitutes a primary challenge. This complexity is further underscored by the fact that conceptualizing AI as persons obscures distinct legal, safety, security, trust, and reliability issues inherent in human-AI relationships~\cite{hoque2024hallmark}.

\textbf{\textcolor{color5}{Challenge-4. Interoperability:}} Seamless user experience requires effective interaction between diverse AI systems, platforms, and tools. AI must integrate smoothly with other technologies, various human workflows, and preferred interfaces (e.g., text, image, voice) to smooth task and context transitions. However, ensuring compatibility across disparate software/hardware, particularly with varying standards and protocols, poses significant technical challenges. Maintaining consistent performance, efficiency, and usability across interconnected AI tools is also a substantial hurdle. This challenge manifests in user experiences of limited controllability in GenAI systems~\cite{sun2024generative}. While specific tools like BrainFax demonstrate efforts to enhance collaborative affordance and bridge systems~\cite{verheijden2023collaborative}, establishing robust, widespread interoperability remains a critical design and engineering goal.

\textbf{\textcolor{color5}{Challenge-5. Social Impact and Equity:}} AI should yield socially responsible and equitable outcomes, promoting diversity while avoiding the reinforcement of inequalities. Key risks include potential job displacement via automation~\cite{weisz2022better}, generating inappropriate or harmful content~\cite{zhang2024mathemyths}, and disrupting creative economies, which may face user objections~\cite{mirowski2023co}. The core challenge, therefore, is ensuring AI benefits society broadly and equitably, actively mitigating harms ranging from human task erasure~\cite{weisz2022better} to economic disruption~\cite{mirowski2023co}. Furthermore, addressing this challenge is compounded by inherent difficulties in reliably assessing societal harm, as noted by Wang et al.~\cite{wang2024farsight}. While human-centered frameworks~\cite{weisz2022better}, harm-aware prototyping~\cite{wang2024farsight}, and inclusive designs~\cite{li2023want} offer valuable pathways, achieving genuine social equity with AI demands ongoing evaluation and profound design shifts.

\textbf{\textcolor{color5}{Challenge-6. Ethical and Bias Concerns:}} AI systems pose ethical issues including generating inaccuracies~\cite{sun2024metawriter}, violating academic integrity~\cite{liu2024coquest}, amplifying dominant views~\cite{sharma2024generative}, and fostering human biases like confirmation bias and over-reliance~\cite{liu2024coquest}. As bias originates from diverse sources (data, models, interaction), addressing these concerns is challenging. Developing inherently unbiased AI and effective feedback mechanisms for bias detection and mitigation is complex, requiring continuous, resource-intensive monitoring and user involvement. While integrating ethical considerations into design shows promise for promoting awareness and practice~\cite{lin2021engaging}, comprehensively resolving these multifaceted issues remains an insufficient, ongoing effort demanding further research.

\textbf{\textcolor{color5}{Challenge-7. Long-term Paradigm Shift:}} A key challenge stems from the limited scope in current discussions concerning AI's impact on co-creation and collaboration models. There is a failure to adequately address the long-term implications of AI's increasing autonomy. This gap results in insufficient consideration of potential significant shifts in the human-machine collaboration paradigm, hindering proactive adaptation and strategy development for future co-creation scenarios. Fu et al.~\cite{fu2024text} highlighted that by changing the language users in interpersonal communication can change how we interact with each others and influence the self-perceptions, thus their developed AI-mediated communication tools could act as catalyst for change. They highlighted several design suggestions including balancing learning and dependency, supporting and augmenting self-expression, including feedback and reflection mechanisms, etc. Liu et al.~\cite{liu2024coquest} pointed out the ideation system may cause human to blindly accept AI in the future, leading to negative impacts in the future.

\begin{figure}[!htbp]
    \includegraphics[width=1\textwidth]{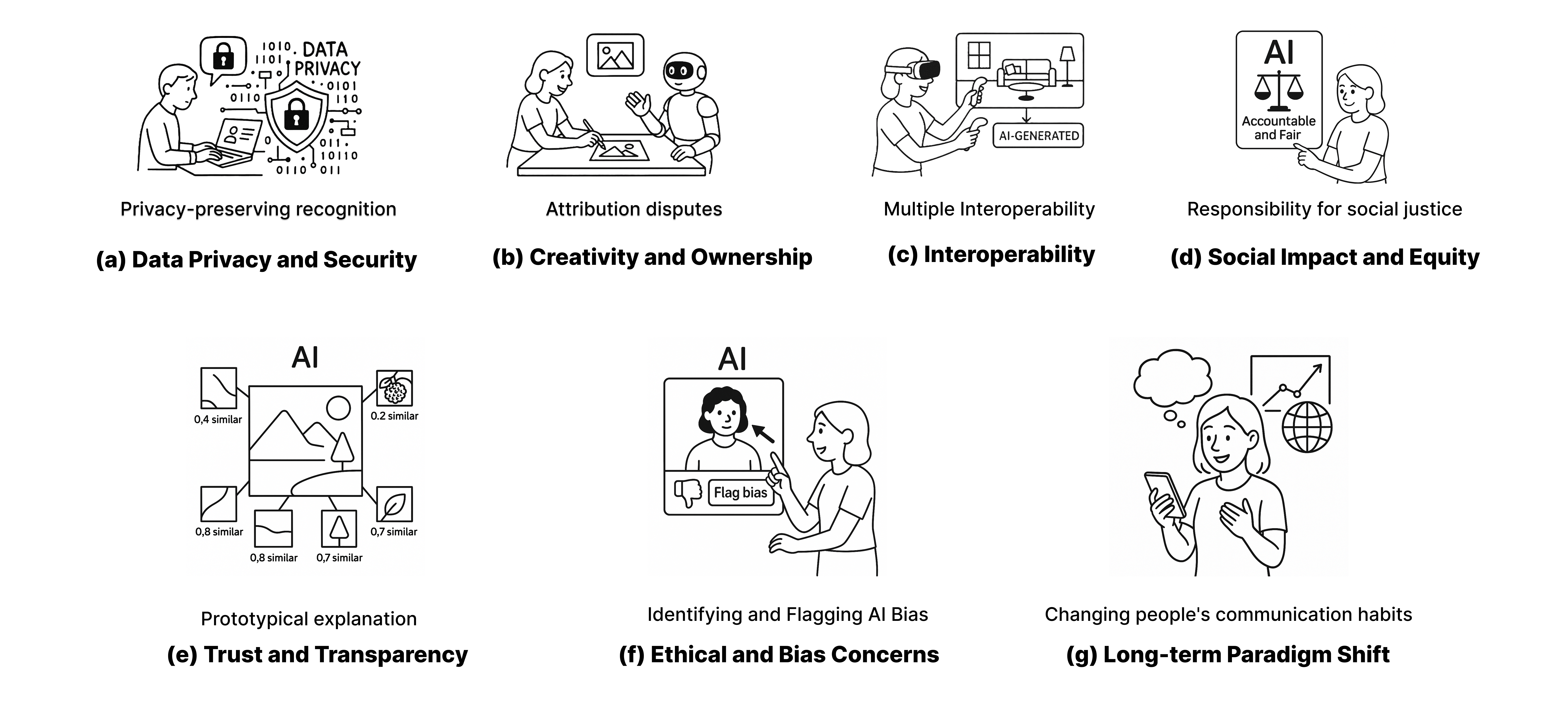}
    \caption{Illustration of design implications and key challenges. The illustrations were generated by ChatGPT and modified by authors.}
\end{figure}

\section{Flow Analysis, Case Analysis and Discussions}\label{sec:findings}

\begin{figure}
    \includegraphics[width=\textwidth]{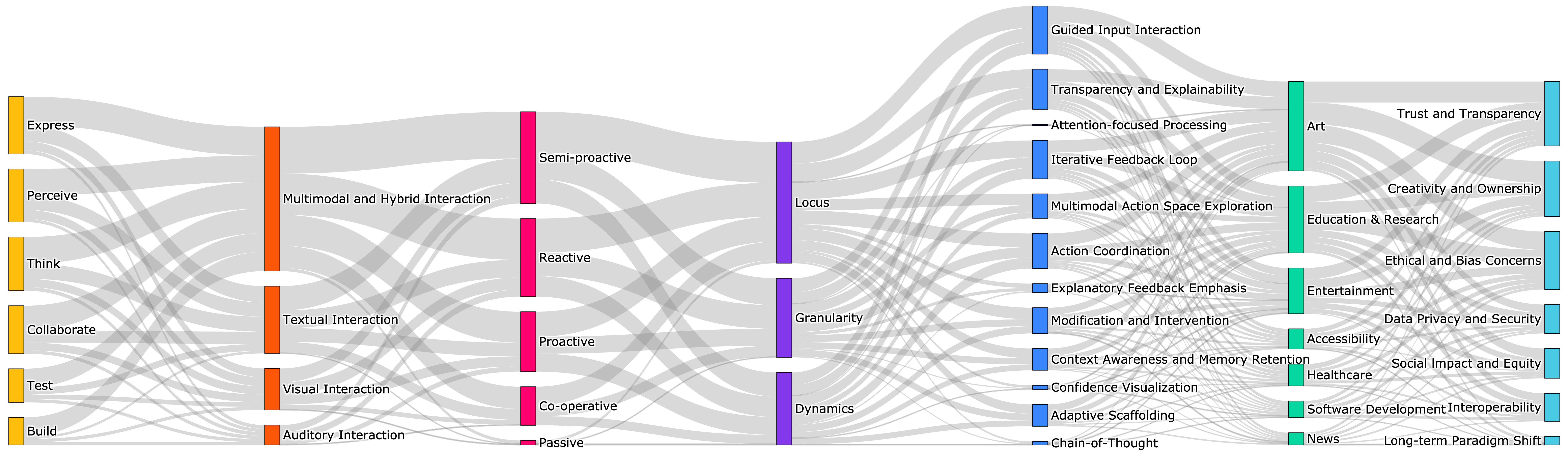}
    \caption{A Sankey diagram summarizing the surveyed literature across critical dimensions we proposed. Flow thickness indicates the frequency of observed connections between concepts. From left to right, the columns represent: \textbf{Interaction Stage}, \textbf{Modality}, \textbf{Agency Patterns}, \textbf{Agency Distribution}, \textbf{Control Mechanisms}, \textbf{Application Domains}, and \textbf{Challenges \& Directions}.}
    \label{fig:co_creation}
\end{figure}

Our literature survey systematically analyzes human-AI co-creation, focusing on agency dynamics. We visualize these findings using a Sankey diagram (Figure~\ref{fig:co_creation}), which maps the flow and relationships across critical dimensions. This structured representation reveals prevalent patterns, interaction dynamics, and potential gaps within co-creative systems, offering concrete application guidelines relevant to CSCW research and design.

\subsection{Node Analysis: Highlighting Core Component Cases}

Node analysis identifies frequently instantiated categories, indicating dominant system configurations. Significant flow through \textbf{``Textual Interaction''} confirms language as the predominant modality. This centrality is evident in systems from academic writing support (e.g., MetaWriter~\cite{sun2024metawriter}) to AI-assisted coding~\cite{wu2022ai,weisz2022better}, where text (natural language or code) forms the primary input and output. While robust NLP remains crucial, the comparative lack of flow through other modalities suggests underexplored design spaces.

High frequency for the \textbf{``Collaborate''} interaction stage signals a strong field emphasis on iterative human-AI partnership. Systems enabling cycles of critique and refinement, like WordCraft~\cite{yuan2022wordcraft} for story editing, or those requiring explicit turn-taking and human validation~\cite{ashktorab2021effects,yang2022ai}, exemplify this stage. This necessitates careful design attention to feedback mechanisms, transparency~\cite{zhu2018explainable,hoque2024hallmark}, and managing negotiated agency~\cite{holter2024deconstructing,lee2024and}.

\textbf{``Co-operative Agency''}, where AI actively partners with the user, constitutes another significant node. Examples include FARSIGHT~\cite{wang2024farsight} exploring AI harms and Zhou et al.'s~\cite{zhou2023beyond} prosthesis personalized systems. This pattern suggests a design trajectory positioning AI as a collaborator, not merely a tool, demanding interfaces that support mutual contribution and shared control.

\subsection{Pathway Analysis: Visualizing Interaction Flow Cases}

Examining dominant pathways through the Sankey diagram (Figure~\ref{fig:co_creation}) reveals recurrent workflows in human-AI co-creation, illuminating the interplay between interaction stages, modalities, agency patterns, control mechanisms, and application domains. These pathways characterize how systems orchestrate collaboration. A prevalent pathway, particularly in writing and software development, centers on textual interaction and co-operative agency: \textbf{Pathway: Perceive/Collaborate/Build $\rightarrow$ Textual Modality $\rightarrow$ Co-operative $\rightarrow$ Feedback Loops/Transparency and Explainability $\rightarrow$ Education \& Research} 

This workflow typically commences with the system perceiving textual input (e.g., data, prompts, code). Collaboration unfolds primarily via the textual modality, progressing through iterative cycles of thinking, expressing, and refining (Collaborate/Build stages). The AI commonly exhibits co-operative agency, functioning as an active partner \cite{mieczkowski2022ai}. Agency distribution often involves a shared locus of control, potentially managed through dynamic allocation negotiated via interaction \cite{holter2024deconstructing, dudley2018review}, with varying granularity depending on the task \cite{zhou2024understanding}. Critical control mechanisms supporting this pathway include iterative feedback loops and transparency/explainability features \cite{hoque2024hallmark, moruzzi2024user} to manage the joint effort. Systems like MetaWriter \cite{sun2024metawriter} (meta-review generation) and AI coding assistants \cite{weisz2022better, wu2022ai} exemplify this trajectory.

Visual co-creation frequently follows a pathway initiated by prompts and emphasizes reactive or co-operative AI roles: \textbf{Pathway: Perceive/Collaborate/Build $\rightarrow$ Visual Modality $\rightarrow$ Reactive/Co-operative $\rightarrow$ Guided Input Interaction/Iterative Feedback Loop $\rightarrow$ Artistic Creation}

In this flow, systems perceive visual data or textual prompts. Interaction proceeds through visual modalities (e.g., image generation interfaces) during the Build and Collaborate stages. AI agency often starts as reactive to the initial prompt \cite{gmeiner2023exploring, lukoff2021design} (AI-centric locus) but may shift towards co-operative during refinement (shared locus) \cite{mieczkowski2022ai}. Control is frequently exerted through guided input (e.g., prompt engineering \cite{lin2024text, tholander2023design}) and iterative refinement cycles \cite{moruzzi2024user}. The locus of control can be dynamic, shifting between human guidance and AI generation \cite{anthraper2024peerconnect, louie2020novice}, often operating at a fine granularity for detailed adjustments \cite{zhou2024understanding}. Examples include text-to-image generation systems \cite{lin2024text} and context-aware image tools like ContextCam \cite{fan2024contextcam}.

Complex co-creation, often involving multiple modalities and leading to sophisticated outputs, necessitates integrated control mechanisms: \textbf{Pathway: Perceive/Collaborate/Build/Test $\rightarrow$ Multimodal Modality $\rightarrow$ Co-operative/Semi-active $\rightarrow$ Guided Input Interaction/Iterative Feedback Loop/Action Coordination $\rightarrow$ Software Development}

This pathway characterizes processes involving diverse inputs perceived by the system, followed by interaction via multimodal or hybrid modalities during Build/Collaborate stages. AI agency patterns are typically co-operative or semi-active \cite{weisz2022better, hegemann2023computational}, often involving a shared or dynamically negotiated locus of control \cite{anthraper2024peerconnect, louie2020novice, dudley2018review}. Effective management relies heavily on integrated control mechanisms, including guided input, action coordination frameworks, and robust feedback loops \cite{zhou2023beyond, hoque2024hallmark, moruzzi2024user}. The granularity of control might be high-level (strategic direction \cite{gu2024data}) or fine-grained (parameter tuning \cite{zhou2024understanding}) depending on the phase. Zhou et al.'s prosthetic personalization system \cite{zhou2023beyond}, involving multimodal interaction (movement affecting form generation) and co-operative agency, fits this pattern, likely leveraging guided input. Similarly, Ayobi et al.'s use of computational notebooks for healthcare co-design \cite{ayobi2023computational} involves multimodal interaction, co-operative agency, and relies on guided input and transparency mechanisms within the notebook structure.

\subsection{Implications from Case Analysis: Guiding Design and Research}

This taxonomy and Sankey analysis (Figure~\ref{fig:co_creation}) yield actionable insights for the CSCW community. Mapping interaction flows reveals prevalent patterns and underexplored areas, such as the sparse use of \textit{``Auditory Interaction''} \cite{kamath2024sound, louie2020novice, sun2024understanding}, signaling innovation opportunities in domains like assistive technology or music co-creation. Observed correlations between \textit{Interaction Stages} (e.g., \textit{``Collaborate''} \cite{holter2024deconstructing}), \textit{AI Agency Patterns} (e.g., \textit{``Proactive''} \cite{hwang2022too} vs. \textit{``Passive''} \cite{suh2021ai}), and \textit{Modalities} \cite{shi2023hci} directly inform \textit{Control Mechanism} design. For instance, frequent \textit{``Collaborate''} stages demand robust \textit{``Iterative Feedback Loops''} \cite{moruzzi2024user, shaer2024ai} and \textit{``Transparency and Explainability''} mechanisms~\cite{hoque2024hallmark, wang2024farsight, zhu2018explainable} for effective partnership management. Systems targeting high AI initiative may require distinct control granularity~\cite{zhou2024understanding} or dynamic agency allocation strategies~\cite{holter2024deconstructing}, unlike systems with primarily reactive or assistive AI.

The framework also helps navigate critical design challenges and enables systematic system comparison. Tracing pathways to \textit{``Ethical and Bias Concerns''} or \textit{``Data Privacy''} nodes  highlights areas demanding rigorous consideration during design and deployment \cite{varanasi2023currently}. For example, \textit{textual interaction} employing \textit{reactive agency} for \textit{personalized news media} \cite{rezk2024agency} surfaces complex ethical questions about filter bubbles, user autonomy, and accountability \cite{sharma2024generative}. Similarly, applications in sensitive domains like \textit{healthcare} \cite{ayobi2023computational, berge2023designing} or \textit{education} \cite{han2024teachers, li2023want} mandate careful balancing of utility against privacy risks and bias mitigation. 

Moreover, the taxonomy provides a structured foundation for comparing diverse co-creation systems. Researchers can accurately situate new contributions \cite{lee2024and}, and practitioners can select tools aligned with specific collaborative goals by evaluating systems--such as FARSIGHT \cite{wang2024farsight}, MetaWriter \cite{sun2024metawriter}, or text-to-image tools \cite{fan2024contextcam}--based on their distinct multidimensional profiles. This integrated analysis offers a dynamic tool for understanding the complex human-AI co-creation landscape, thereby fostering principled design innovation within CSCW.

\subsection{Long-term Implications of AI's Impact on Co-Creation Models}
\label{sec:implications}

Our synthesis indicates AI autonomy drives long-term shifts~\cite{fu2024text} theoretically, technically and socio-economically. Theoretically, AI's evolution from functional tool~\cite{shneiderman1997direct} to autonomous, proactive partner~\cite{trajkova2024exploring,suh2021ai,aagerfalk2020artificial} challenges human-centric assumptions about creativity and requires new collaboration theories beyond mixed-initiative~\cite{horvitz1999principles}, potentially involving co-agency or emergent interaction structures~\cite{wu2021ai,holter2024deconstructing}. This evolution may also profoundly alter human cognition and self-perception~\cite{fu2024text}, risking deskilling~\cite{zhang2022you} or over-reliance~\cite{liu2024coquest,kamath2024sound,liu2021ai}, yet simultaneously offering strong augmentation to human~\cite{dhillon2024shaping}. Investigating how to foster beneficial cognitive relationships with autonomous AI is thus crucial for CSCW community~\cite{fu2024text}.

Technically, increasing AI autonomy necessitates collaborative models extending beyond supervisory control~\cite{holter2024deconstructing} towards peer-like or even AI-coordinated dynamics. This demands novel technical solutions and interaction mechanisms for negotiating agency dynamically~\cite{holter2024deconstructing}, ensuring accountability for autonomous actions~\cite{wang2024farsight}, providing explainability that clarifies AI reasoning and agency~\cite{zhu2018explainable,weisz2024design}, managing complex power dynamics~\cite{li2023beyond}, enabling graceful human intervention, and fostering calibrated trust~\cite{lee2004trust,liu2021ai}. Addressing potential conflicts between humans and autonomous AI~\cite{hodhod2016closing} and ensuring interoperability within complex human-AI ecosystems~\cite{wu2021ai} represent significant steps for future systems.

Societally, the integration of autonomous AI into co-creation could significantly restructure creative industries and labor markets~\cite{gmeiner2023exploring,berge2023designing,mieczkowski2022ai}, raising critical questions about the future of creative work, including intellectual property~\cite{kahveci2023attribution}, fair compensation~\cite{kadoma2024role}, and evolving skill requirements~\cite{zhang2022you}. Furthermore, AI autonomy intensities ethical challenges concerning bias amplification~\cite{sharma2024investigating}, fairness~\cite{verma2019weapons}, transparency~\cite{hoque2024hallmark}, accountability for harms~\cite{wang2024farsight}, and overall societal responsibility~\cite{salatino2025influence}. Proactive development of adaptable governance frameworks and commitment to value-aligned design practices~\cite{varanasi2023currently,lin2021engaging} are essential for navigating these complexities.

\subsection{Connection to Agency Theories, Broad Fields and Limitations}
\label{sec:theory_correlation}

This review's findings offer empirical grounding for established agency theories, inform broad AI/ML fields and highlight areas for future theoretical development.

The observed diversity in agency configurations (Sections~\ref{sec:agency_patterns}, \ref{sec:agency_distribution}) provides concrete empirical support for theories framing agency as distributed, situated, and dynamically negotiated \cite{suchman1987plans,wu2021ai,holter2024deconstructing}. Our synthesis shows systems instantiating varied distributed agency arrangements \cite{holter2024deconstructing}, from human-centric \cite{dhillon2024shaping} to shared partnerships \cite{zheng2024charting}. Furthermore, the comprehensive catalog of control mechanisms (Section~\ref{sec:control_mechanisms}) illuminates the crucial operational layer linking user intention (agency) to system execution. This aligns with theories distinguishing agency from control \cite{pacherie2007sense, limerick2014experience}, detailing the practical interface mechanisms (e.g., `Guided Input' \cite{lin2024text}, `Intervention' \cite{hoque2024hallmark}) that shape the user's experience of agency \cite{limerick2014experience}. The emergence of co-operative agency patterns (Section~\ref{sec:agency_patterns}), particularly when supported by mechanisms enabling `Action Coordination' (Section~\ref{sec:control_mechanisms}), invites consideration of human-AI dyads through the lens of group agency \cite{list2011group}. 

This operational perspective underscores that raw AI capabilities~\cite{sharma2024investigating} (e.g., predictive accuracy, generative power) developed in ML are necessary but insufficient for effective co-creation~\cite{mitelut2024position}. Mitelut et al.~\cite{mitelut2024position} proposed that alignment need to optimize for agency preservation, which our paper from the human-centric perspective provide the detailed control mechanisms for agency preservation. Researchers also proposed data agency theory~\cite{ajmani2024data}, a precise theory of justice to evaluate and improve current consent procedures used in AI applications. Our work proposed the control mechanisms at the next step of their evaluation.

Despite these connections, our review also underscores theoretical areas requiring further development for this context. First, existing agency theories offer limited frameworks for modeling the dynamic transitions in roles and control observed in our findings (Sections 5.2, 9.2). This limitation implies that our categorization of agency patterns (Section 5.1) and interaction pathways (Section 9.2) may not fully capture the moment-to-moment negotiations occurring in practice. Future theoretical work could develop more fine-grained models, perhaps drawing on process theories~\cite{forbus1984qualitative} or computational approaches~\cite{abraham2009computational}. Second, a significant theoretical gap persists in reconciling the AI's designed, computational nature with the user's subjective experience and attribution of agency~\cite{limerick2014experience,sundar2020rise}. This impacts the interpretation of our findings on trust, ethics and ownership (Section 8), as the discrepancy between how AI functions (Section 6) and how users perceive its intentionality~\cite{searle1980minds,bennett2023does} influences responsibility alignment and bias perception~\cite{sharma2024generative,varanasi2023currently} in ways our current framework only partially addresses. Third, the applicability of general agency theories (Section 7) is less clear for the open-ended, emergent goals characteristic of many co-creative applications (e.g., artwork). Our analysis highlights diverse practices, but the theoretical underpinnings of agency in highly exploratory versus goal-directed co-creation need further distinction.


\section{Conclusion}
This scoping review systematically mapped the HCI literature on human-AI co-creation through the lens of agency. Analyzing 134 papers from premier HCI venues, we proposed an analytical framework integrating context, agency patterns/distribution, control mechanisms, and embedded them in practical applications. We cataloged operational control mechanisms detailing agency implementation. Findings highlight diverse agency configurations and underscore the critical role of specific control mechanisms (e.g., transparency, feedback loops) in shaping collaboration, user experience, and trust. The synthesis offers grounded implications for CSCW, guiding the design of future co-creative systems concerning ethical, societal and long-term implications.

\begin{acks}
We sincerely thank all reviewers for providing their valuable feedback. We thank Yongquan Hu for providing valuable help during the work. This work was supported by the Natural Science Foundation of China under Grant No. 62472243 and 62132010. This work was also supported by the Deng Feng Fund.
\end{acks}
\bibliographystyle{ACM-Reference-Format}
\bibliography{sample-base}

\appendix 
\section{PRISMA-ScR Guided Literature Review Process and Justification}\label{sec:prisma}

\begin{figure}
    \centering
    \includegraphics[width=0.8\textwidth]{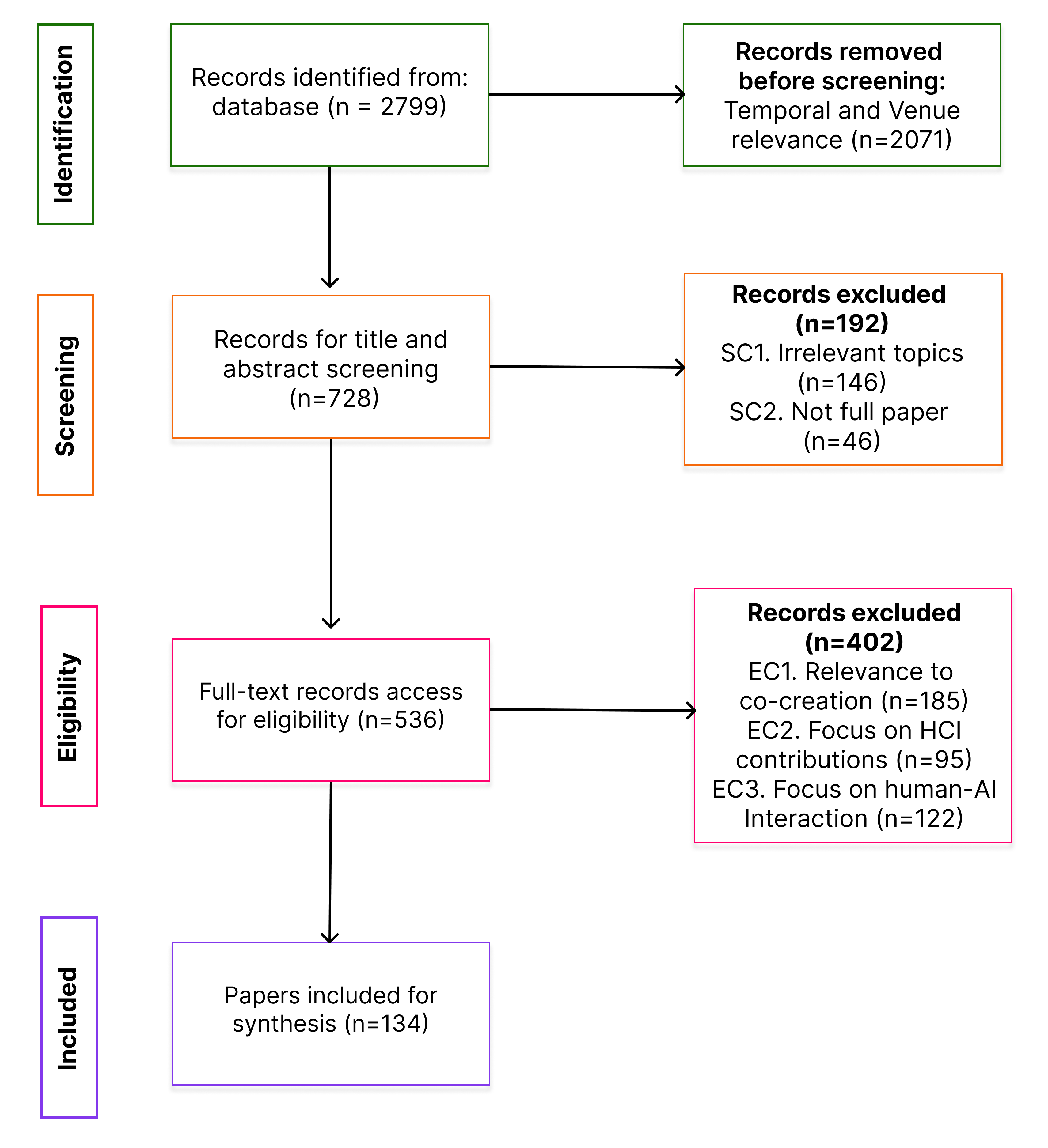}
    \caption{The PRISMA-ScR guided process of our paper selections for the review.}
    \label{fig:prisma}
\end{figure}

Our literature review followed PRISMA-ScR guideline, and the whole process is outlined in Figure~\ref{fig:prisma}.

\textbf{Literature Search Details} Our query string contained the uppercase and lowercase version of (``co-creation'' OR ``co-writing'' OR ``co-drawing'' OR ``co-design'') AND (``agency''), applied to the title and content. The terms ``co-creation'', ``co-writing'', ``co-drawing'', ``co-design'' were iteratively optimized and selected to represent the most common and explicit descriptors for human-AI collaborative activities studied HCI. They cover a range of creative and design-oriented tasks. We have attempted to add ``co-dancing'', ``co-making'' and others, which resulted in no additional papers. For ``agency'', this is the core theoretical construct under investigation, and we want to make sure the paper explicitly mentioned agency. We also attempted to add similar words like ``autonomy'', but resulted in no increase of eligible papers. \textbf{We selected conferences including CHI, UIST, CSCW, Ubicomp, IUI, DIS manually when conducting the searching to scope the search in top-tier HCI conferences. }

\textbf{Exclusion Criteria Definitions} The following provides clarification on the exclusion criteria applied during the full-text coding. 

\textit{Screening Criteria (Applied to Titles \& Abstracts)}

\textbf{SC1-Irrelevant Topic:}  Papers whose titles or abstracts clearly indicated a primary focus outside the scope of human-AI co-creation and agency were excluded. This initial thematic filter was crucial for efficiency, removing papers clearly unrelated to the review's core concentration on co-creation and agency before committing to the full-text analysis. Exampls of excluded records included papers solely on AI algorithms, non-AI HCI studies, general collaboration theories.

\textbf{SC2-Not Full Paper:} Records identified as not being peer-reviewed full papers appropriate for this review were excluded. This criterion ensured the review focused on substantial works typical of the targeted top-tier venues. Excluding short and non-peer-reviewed formats helps maintain a consistent depth and quality of evidence across the analyzed corpus, aligning with the goal of synthesizing established research findings. Examples of excluded records include workshop summaries, demo descriptions, dissertations.

\textit{Eligibility Criteria (Applied to Full Texts)}

\textbf{EC1-Relevance to Co-creation:} Papers not focused on collaborative or co-creative processes involving both humans and AI systems within an HCI context were excluded. This criterion confirmed that the core subject of the paper involved human-AI co-creation rather than just any human-AI interaction. Examples of excluded papers include AI algorithm optimization studies, autonomous navigation without co-creation tools, non-collaborative AI tools.

\textbf{EC2-Focus on Concrete HCI Contributions:} Papers discussing agency or co-creation primarily at a conceptual, philosophical, or ethical level without detailing or evaluating specific HCI techniques, interaction designs, or implemented systems were excluded. This criterion was vital for ensuring that our review could meet its objective of cataloging and analyzing the operational aspects of agency,  specifically, the concrete control mechanisms and interaction designs used in practice. By focusing on papers with HCI contributions, our review could synthesize actionable design knowledge and identify patterns in implemented systems, directly addressing the identified gap regarding practical operationalization of agency. Examples of excluded papers involve high-level framework proposals.

\textbf{EC3-Focus on Human-AI Interaction:} Papers where the interaction studied or designed involved only humans interacting with other humans or only AI agents interacting with other AI agents were excluded. It ensured that the analysis focused on the unique interaction phenomena, agency distributions, and control challenges arising specifically at the interface between human users and AI systems, which is the central locus of investigation for this review. Studies of purely human or purely AI systems, while potentially relevant to collaboration broadly, fall outside this specific scope. Examples of excluded papers include studies on human-human collaboration tools or AI-AI multi-agent systems.

\section{Coding Criteria and Theoretical Foundations}\label{sec:coding}

We adopted the following theories and criteria when coding the papers. 

\begin{itemize} 
    \item \textit{Research Methods}: Categorized into quantitative, qualitative, theoretical/evaluation framework, mixed methods, and user-centered research, following Bi et al.~\cite{bi2019characterizing}.  
    \item \textit{Research Types/Contributions}: Classified as comparison study, concept generation, design, system implementation, or interview\footnote{\url{https://ueberproduct.de/en/4-types-of-research/}}.
    \item \textit{Interaction Stage (one dimension of Context):} Coded based on the six stages (perceive, think, express, collaborate, build, test) derived from Wu et al.~\cite{wu2021ai} and utilized by Zhang et al.~\cite{zhang2023towards}. 
    \item \textit{Interaction modality (one dimension of Context):} Classified into Textual, Visual, Auditory, and Multimodal/Hybrid, informed by prior work~\cite{JustThinkAI2024CoCreation,shi2023hci} and observations in the literature.
    \item \textit{Levels of Agency:} Categorized based on Rammert et al.'s framework~\cite{mieczkowski2022ai}, into passive, semi-active, reactive, proactive, cooperative.
    \item \textit{Agency Distribution:} Analyzed using the three dimensions extending Holter et al.~\cite{holter2024deconstructing}: locus of control, control dynamics (static/dynamic allocation), and granularity of control. Here the agency distribution included the ones the studies in the literature explicitly covered and the ones in their design / implementation.
    \item \textit{Control Mechanisms:} we organized the control mechanisms according to IPOF model stages (Input, Action, Output, Feedback)~\cite{wiener2019cybernetics}, and then devised the detailed control mechanisms through manually synthesizing from the literature. 
    \item \textit{Applications:} we organized the applications through identifying and aggregating potential applications from the the literature. Here the applications encompassed all targeted applications the authors claimed. We did not directly adopt Wu et al.'s work~\cite{wu2021ai} because their classification was at an abstract level and faces limitation in guiding actionable designs.
    \item \textit{Challenges:} We manually coded and organized the challenges from the literature. Here the challenges denoted all stated challenges or directions in the literature.
    
\end{itemize}

\section{The Taxonomy of Papers}

\begin{table}
\centering 
\caption{The interaction stages the papers in our surveys are based on.}
\label{tbl:interaction_phase}
\begin{tabularx}{\textwidth}{p{0.3\textwidth}p{0.7\textwidth}}
\toprule
\textbf{Interaction Stage} & \textbf{Papers} \\ \midrule 
Perceive & ~\cite{dhillon2024shaping,lin2024text,tholander2023design,moruzzi2024user,hoque2024hallmark,sun2024generative,weisz2022better,kamath2024sound,wang2024farsight,berge2023designing,sun2024metawriter,cremaschi2024steampunk,gebreegziabher2023patat,sharma2024generative,benjamin2023entoptic,fu2024text,fan2024contextcam,verheijden2023collaborative,liu2024ai,claisse2024understanding,suh2024luminate,osone2021buncho,louie2020novice,mirowski2023co,zheng2024charting,zhou2023beyond,chang2024co,ayobi2023computational,bircanin2021including,lin2021engaging,gmeiner2023exploring,kadoma2024role,suh2021ai,newman2024want,anthraper2024peerconnect,zhang2024mathemyths,de2024bias,wang2023treat,trajkova2024exploring,lukoff2021design,rezk2024agency,praveena2023exploring,han2024teachers,bauer2024musictraces,zhang2022get,weisz2024design,shaer2024ai,wan2024metamorpheus,elagroudy2024transforming,kang2024design,brand2023envisioning,liu2024he,varanasi2024saharaline,guo2024exploring,lu2022bridging,vear2024jess+,lee2024open,lyu2024preliminary,wallace2022asynchronous,shi2020emog,inie2023designing,almeda2024prompting,lazar2018making,mahdavi2024ai,arakawa2023catalyst,belghith2024testing,mccormack2019silent,jones2023embodying,albaugh2023augmented,wang2024critical,reig2023supporting,shibani2024untangling,simpson2023rethinking,soler2024arcadia,gebreegziabher2023patat,sharma2023takes,hegemann2023computational,zong2024umwelt,jones2024customization,ashktorab2021effects,park2023importance,lee2024design,von2023designing,leong2024putting,yang2023pair,choi2023creator,neate2019empowering,bala2023towards,wu2022ai,durrant2011automics,billewicz2018silly,shelby2024generative,milton2023see,zhang2022storybuddy,li2023understanding,maeng2022designing,khaled2013design,chung2022artistic,liu2022opal,dang2022beyond,jain2023front,hong2022scholastic,li2023beyond,zhang2023towards,cheon2017configuring,wang2024lave,louie2022expressive,lawton2023drawing,yuan2022wordcraft,hodhod2016closing,bako2023user} \\ \hline 
Think & ~\cite{dhillon2024shaping,lin2024text,tholander2023design,moruzzi2024user,hoque2024hallmark,sun2024generative,weisz2022better,kamath2024sound,wang2024farsight,berge2023designing,sun2024metawriter,cremaschi2024steampunk,gebreegziabher2023patat,benjamin2023entoptic,fu2024text,fan2024contextcam,verheijden2023collaborative,liu2024ai,claisse2024understanding,suh2024luminate,osone2021buncho,louie2020novice,sun2024understanding,mirowski2023co,zheng2024charting,zhou2023beyond,chang2024co,ayobi2023computational,li2023want,lin2021engaging,gmeiner2023exploring,kadoma2024role,suh2021ai,anthraper2024peerconnect,zhang2024mathemyths,de2024bias,wang2023treat,trajkova2024exploring,hwang2022too,lukoff2021design,rezk2024agency,praveena2023exploring,han2024teachers,bauer2024musictraces,zhang2022get,weisz2024design,shaer2024ai,wan2024metamorpheus,elagroudy2024transforming,kang2024design,brand2023envisioning,liu2024he,varanasi2024saharaline,guo2024exploring,lu2022bridging,vear2024jess+,lee2024open,chang2024dark,lyu2024preliminary,shi2020emog,inie2023designing,almeda2024prompting,lazar2018making,arakawa2023catalyst,belghith2024testing,mccormack2019silent,jones2023embodying,albaugh2023augmented,wang2024critical,shibani2024untangling,simpson2023rethinking,soler2024arcadia,gebreegziabher2023patat,sharma2023takes,hegemann2023computational,zong2024umwelt,jones2024customization,ashktorab2021effects,lee2024design,von2023designing,leong2024putting,yang2023pair,choi2023creator,neate2019empowering,bala2023towards,wu2022ai,billewicz2018silly,shelby2024generative,varanasi2023currently,milton2023see,zhang2022storybuddy,li2023understanding,marquez2016embodied,maeng2022designing,khaled2013design,chung2022artistic,liu2022opal,dang2022beyond,zhang2023visar,jain2023front,hong2022scholastic,li2023beyond,zhang2023towards,cheon2017configuring,wang2024lave,louie2022expressive,lawton2023drawing,yuan2022wordcraft,hodhod2016closing,bako2023user} \\ \hline 
Express & ~\cite{dhillon2024shaping,lin2024text,tholander2023design,moruzzi2024user,hoque2024hallmark,sun2024generative,weisz2022better,kamath2024sound,wang2024farsight,berge2023designing,sun2024metawriter,cremaschi2024steampunkv,gebreegziabher2023patat,sharma2024generative,benjamin2023entoptic,fu2024text,fan2024contextcam,verheijden2023collaborative,liu2024ai,claisse2024understanding,suh2024luminate,osone2021buncho,louie2020novice,sun2024understanding,mirowski2023co,zheng2024charting,zhou2023beyond,chang2024co,ayobi2023computational,bircanin2021including,li2023want,lin2021engaging,gmeiner2023exploring,kadoma2024role,suh2021ai,newman2024want,anthraper2024peerconnect,zhang2024mathemyths,de2024bias,wang2023treat,trajkova2024exploring,hwang2022too,lukoff2021design,rezk2024agency,praveena2023exploring,han2024teachers,bauer2024musictraces,zhang2022get,weisz2024design,shaer2024ai,wan2024metamorpheus,elagroudy2024transforming,kang2024design,brand2023envisioning,liu2024he,varanasi2024saharaline,guo2024exploring,lu2022bridging,vear2024jess+,lee2024open,chang2024dark,lyu2024preliminary,wallace2022asynchronous,shi2020emog,almeda2024prompting,lazar2018making,mahdavi2024ai,arakawa2023catalyst,belghith2024testing,mccormack2019silent,jones2023embodying,albaugh2023augmented,wang2024critical,reig2023supporting,shibani2024untangling,simpson2023rethinking,soler2024arcadia,gebreegziabher2023patat,sharma2023takes,hegemann2023computational,zong2024umwelt,jones2024customization,ashktorab2021effects,park2023importance,lee2024design,von2023designing,chakrabarty2024art,leong2024putting,yang2023pair,choi2023creator,neate2019empowering,bala2023towards,wu2022ai,durrant2011automics,billewicz2018silly,shelby2024generative,varanasi2023currently,milton2023see,zhang2022storybuddy,li2023understanding,marquez2016embodied,maeng2022designing,khaled2013design,chung2022artistic,liu2022opal,dang2022beyond,zhang2023visar,jain2023front,hong2022scholastic,li2023beyond,zhang2023towards,cheon2017configuring,wang2024lave,louie2022expressive,lawton2023drawing,yuan2022wordcraft,hodhod2016closing,bako2023user} \\ \hline 
Collaborate & ~\cite{dhillon2024shaping,lin2024text,tholander2023design,moruzzi2024user,hoque2024hallmark,sun2024generative,kamath2024sound,wang2024farsight,berge2023designing,sun2024metawriter,cremaschi2024steampunk,gebreegziabher2023patat,sharma2024generative,fu2024text,fan2024contextcam,verheijden2023collaborative,liu2024ai,claisse2024understanding,suh2024luminate,osone2021buncho,louie2020novice,sun2024understanding,mirowski2023co,zheng2024charting,zhou2023beyond,chang2024co,ayobi2023computational,bircanin2021including,li2023want,lin2021engaging,gmeiner2023exploring,suh2021ai,newman2024want,anthraper2024peerconnect,zhang2024mathemyths,wang2023treat,trajkova2024exploring,hwang2022too,lukoff2021design,rezk2024agency,praveena2023exploring,bauer2024musictraces,zhang2022get,weisz2024design,shaer2024ai,wan2024metamorpheus,elagroudy2024transforming,kang2024design,brand2023envisioning,liu2024he,varanasi2024saharaline,guo2024exploring,capel2023human,vear2024jess+,wallace2022asynchronous,shi2020emog,lazar2018making,mahdavi2024ai,mccormack2019silent,albaugh2023augmented,wang2024critical,reig2023supporting,shibani2024untangling,gebreegziabher2023patat,sharma2023takes,hegemann2023computational,ashktorab2021effects,park2023importance,yang2023pair,choi2023creator,neate2019empowering,bala2023towards,wu2022ai,durrant2011automics,billewicz2018silly,shelby2024generative,varanasi2023currently,zhang2022storybuddy,li2023understanding,marquez2016embodied,khaled2013design,chung2022artistic,dang2022beyond,zhang2023visar,hong2022scholastic,zhang2023towards,wang2024lave,louie2022expressive,lawton2023drawing,yuan2022wordcraft,hodhod2016closing,bako2023user} \\ \hline 
Build & ~\cite{dhillon2024shaping,hoque2024hallmark,sun2024generative,weisz2022better,fan2024contextcam,verheijden2023collaborative,claisse2024understanding,osone2021buncho,louie2020novice,zhou2023beyond,ayobi2023computational,bircanin2021including,gmeiner2023exploring,suh2021ai,newman2024want,hwang2022too,han2024teachers,bauer2024musictraces,weisz2024design,wan2024metamorpheus,elagroudy2024transforming,brand2023envisioning,liu2024he,lu2022bridging,lee2024open,lyu2024preliminary,wallace2022asynchronous,shi2020emog,lazar2018making,arakawa2023catalyst,jones2023embodying,reig2023supporting,simpson2023rethinking,sharma2023takes,hegemann2023computational,zong2024umwelt,von2023designing,chakrabarty2024art,choi2023creator,neate2019empowering,durrant2011automics,shelby2024generative,varanasi2023currently,zhang2022storybuddy,khaled2013design,chung2022artistic,liu2022opal,dang2022beyond,li2023beyond,zhang2023towards,louie2022expressive,lawton2023drawing,bako2023user} \\ \hline 
Test & ~\cite{tholander2023design,hoque2024hallmark,sun2024generative,weisz2022better,kamath2024sound,berge2023designing,cremaschi2024steampunk,sharma2024generative,fu2024text,fan2024contextcam,verheijden2023collaborative,liu2024ai,claisse2024understanding,louie2020novice,mirowski2023co,zheng2024charting,zhou2023beyond,ayobi2023computational,lin2021engaging,gmeiner2023exploring,suh2021ai,newman2024want,anthraper2024peerconnect,zhang2024mathemyths,trajkova2024exploring,lukoff2021design,rezk2024agency,praveena2023exploring,han2024teachers,bauer2024musictraces,weisz2024design,shaer2024ai,elagroudy2024transforming,brand2023envisioning,varanasi2024saharaline,lu2022bridging,wallace2022asynchronous,almeda2024prompting,mahdavi2024ai,mccormack2019silent,jones2023embodying,wang2024critical,simpson2023rethinking,sharma2023takes,hegemann2023computational,jones2024customization,von2023designing,chakrabarty2024art,leong2024putting,yang2023pair,choi2023creator,neate2019empowering,wu2022ai,durrant2011automics,billewicz2018silly,shelby2024generative,varanasi2023currently,zhang2022storybuddy,maeng2022designing,khaled2013design,chung2022artistic,liu2022opal,jain2023front,louie2022expressive,bako2023user} \\ \bottomrule
\end{tabularx}
\end{table}

\begin{table}
\centering
\caption{The modalities the papers in our surveys are based on.}
\label{tbl:context}
\begin{tabularx}{\textwidth}{p{0.3\textwidth}p{0.7\textwidth}}
\toprule
\textbf{Modality} & \textbf{Papers} \\ \midrule 
textual interaction & \cite{dhillon2024shaping,hoque2024hallmark,weisz2022better,wang2024farsight,sun2024metawriter,jones2023embodying,cremaschi2024steampunk,gebreegziabher2023patat,sharma2024generative,fu2024text,liu2024ai,suh2024luminate,osone2021buncho,mirowski2023co,kadoma2024role,shaer2024ai,kang2024design,guo2024exploring,inie2023designing,almeda2024prompting,arakawa2023catalyst,belghith2024testing,shibani2024untangling,gebreegziabher2023patat,sharma2023takes,jones2024customization,ashktorab2021effects,park2023importance,lee2024design,chakrabarty2024art,leong2024putting,sharma2024generative,neate2019empowering,wu2022ai,fu2024text,maeng2022designing,dang2022beyond,weisz2022better,yuan2022wordcraft} \\ \hline 
visual interaction & \cite{lin2024text,benjamin2023entoptic,verheijden2023collaborative,de2024bias,lukoff2021design,praveena2023exploring,zhang2022get,wan2024metamorpheus,shi2020emog,lazar2018making,mahdavi2024ai,wang2024critical,simpson2023rethinking,hegemann2023computational,von2023designing,choi2023creator,durrant2011automics,shelby2024generative,milton2023see,li2023understanding,lawton2023drawing,hodhod2016closing,bako2023user} \\ \hline 
auditory interaction & \cite{kamath2024sound,louie2020novice,sun2024understanding,suh2021ai,zhang2024mathemyths,kamath2024sound,jain2023front,louie2022expressive} \\ \hline
multimodal and hybrid interaction & \cite{tholander2023design,moruzzi2024user,sun2024generative,berge2023designing,vear2024jess+,fan2024contextcam,claisse2024understanding,zheng2024charting,zhou2023beyond,chang2024co,ayobi2023computational,bircanin2021including,li2023want,lin2021engaging,mucha2020co,gmeiner2023exploring,newman2024want,anthraper2024peerconnect,wang2023treat,trajkova2024exploring,hwang2022too,rezk2024agency,han2024teachers,bauer2024musictraces,weisz2024design,elagroudy2024transforming,brand2023envisioning,liu2024he,varanasi2024saharaline,lu2022bridging,capel2023human,vear2024jess+,lee2024open,chang2024dark,lyu2024preliminary,wallace2022asynchronous,mccormack2019silent,jones2023embodying,albaugh2023augmented,reig2023supporting,soler2024arcadia,zong2024umwelt,yang2023pair,bala2023towards,billewicz2018silly,varanasi2023currently,zhang2022storybuddy,marquez2016embodied,khaled2013design,chung2022artistic,liu2022opal,zhang2023visar,hong2022scholastic,li2023beyond,zhang2023towards,cheon2017configuring,wang2024lave} \\ \bottomrule
\end{tabularx}
\end{table}

\begin{table}
\centering
\caption{The agency patterns the papers in our surveys included.}
\label{tbl:agency}
\begin{tabularx}{\textwidth}{p{0.15\textwidth}p{0.85\textwidth}}
\toprule
\textbf{Agency Pattern} & \textbf{Papers} \\ \midrule 
Passive & \cite{tholander2023design,moruzzi2024user,weisz2022better,cremaschi2024steampunk,zheng2024charting,hwang2022too,lukoff2021design,rezk2024agency} \\ \hline 
Reactive & \cite{dhillon2024shaping,lin2024text,tholander2023design,moruzzi2024user,hoque2024hallmark,sun2024generative,weisz2022better,kamath2024sound,wang2024farsight,berge2023designing,sun2024metawriter,cremaschi2024steampunk,gebreegziabher2023patat,sharma2024generative,benjamin2023entoptic,fu2024text,fan2024contextcam,verheijden2023collaborative,liu2024ai,suh2024luminate,osone2021buncho,louie2020novice,sun2024understanding,mirowski2023co,zheng2024charting,zhou2023beyond,bircanin2021including,li2023want,gmeiner2023exploring,kadoma2024role,suh2021ai,newman2024want,anthraper2024peerconnect,zhang2024mathemyths,de2024bias,wang2023treat,trajkova2024exploring,hwang2022too,lukoff2021design,rezk2024agency,praveena2023exploring,han2024teachers,bauer2024musictraces,zhang2022get,weisz2024design,albaugh2023augmented,wan2024metamorpheus,kang2024design,brand2023envisioning,liu2024he,varanasi2024saharaline,guo2024exploring,vear2024jess+,lee2024open,chang2024dark,lyu2024preliminary,wallace2022asynchronous,shi2020emog,lazar2018making,mahdavi2024ai,arakawa2023catalyst,belghith2024testing,mccormack2019silent,jones2023embodying,albaugh2023augmented,wang2024critical,reig2023supporting,shibani2024untangling,simpson2023rethinking,soler2024arcadia,hegemann2023computational,ashktorab2021effects,von2023designing,chakrabarty2024art,leong2024putting,yang2023pair,choi2023creator,fu2024text,durrant2011automics,billewicz2018silly,shelby2024generative,varanasi2023currently,milton2023see,zhang2022storybuddy,li2023understanding,marquez2016embodied,maeng2022designing,khaled2013design,chung2022artistic,liu2022opal,dang2022beyond,zhang2023visar,jain2023front,hong2022scholastic,zhang2023towards,cheon2017configuring,wang2024lave,louie2022expressive,lawton2023drawing,yuan2022wordcraft,hodhod2016closing,bako2023user} \\ \hline 
Semi-proactive & \cite{dhillon2024shaping,lin2024text,tholander2023design,moruzzi2024user,hoque2024hallmark,sun2024generative,weisz2022better,kamath2024sound,wang2024farsight,berge2023designing,sun2024metawriter,cremaschi2024steampunk,gebreegziabher2023patat,sharma2024generative,benjamin2023entoptic,fu2024text,fan2024contextcam,verheijden2023collaborative,liu2024ai,claisse2024understanding,suh2024luminate,osone2021buncho,louie2020novice,sun2024understanding,mirowski2023co,zheng2024charting,zhou2023beyond,chang2024co,ayobi2023computational,bircanin2021including,li2023want,lin2021engaging,mucha2020co,gmeiner2023exploring,kadoma2024role,suh2021ai,newman2024want,anthraper2024peerconnect,zhang2024mathemyths,de2024bias,wang2023treat,trajkova2024exploring,hwang2022too,lukoff2021design,rezk2024agency,praveena2023exploring,han2024teachers,bauer2024musictraces,zhang2022get,weisz2024design,albaugh2023augmented,wan2024metamorpheus,kang2024design,brand2023envisioning,liu2024he,varanasi2024saharaline,guo2024exploring,lu2022bridging,capel2023human,vear2024jess+,lee2024open,chang2024dark,lyu2024preliminary,wallace2022asynchronous,shi2020emog,inie2023designing,almeda2024prompting,lazar2018making,mahdavi2024ai,arakawa2023catalyst,belghith2024testing,mccormack2019silent,jones2023embodying,albaugh2023augmented,wang2024critical,reig2023supporting,shibani2024untangling,simpson2023rethinking,soler2024arcadia,gebreegziabher2023patat,sharma2023takes,hegemann2023computational,zong2024umwelt,jones2024customization,ashktorab2021effects,park2023importance,lee2024design,von2023designing,chakrabarty2024art,leong2024putting,yang2023pair,sharma2024generative,choi2023creator,neate2019empowering,bala2023towards,wu2022ai,fu2024text,durrant2011automics,billewicz2018silly,shelby2024generative,varanasi2023currently,milton2023see,zhang2022storybuddy,li2023understanding,marquez2016embodied,maeng2022designing,khaled2013design,chung2022artistic,liu2022opal,dang2022beyond,zhang2023visar,jain2023front,hong2022scholastic,zhang2023towards,cheon2017configuring,wang2024lave,louie2022expressive,lawton2023drawing,yuan2022wordcraft,bako2023user} \\ \hline
Proactive & \cite{dhillon2024shaping,moruzzi2024user,sun2024generative,weisz2022better,kamath2024sound,wang2024farsight,berge2023designing,cremaschi2024steampunk,gebreegziabher2023patat,benjamin2023entoptic,fan2024contextcam,claisse2024understanding,suh2024luminate,osone2021buncho,louie2020novice,sun2024understanding,zheng2024charting,zhou2023beyond,chang2024co,ayobi2023computational,bircanin2021including,li2023want,lin2021engaging,mucha2020co,gmeiner2023exploring,kadoma2024role,newman2024want,wang2023treat,trajkova2024exploring,hwang2022too,lukoff2021design,rezk2024agency,praveena2023exploring,han2024teachers,wan2024metamorpheus,kang2024design,brand2023envisioning,lu2022bridging,capel2023human,vear2024jess+,lee2024open,chang2024dark,lyu2024preliminary,lazar2018making,mahdavi2024ai,wang2024critical,reig2023supporting,shibani2024untangling,simpson2023rethinking,soler2024arcadia,gebreegziabher2023patat,sharma2023takes,zong2024umwelt,jones2024customization,park2023importance,lee2024design,sharma2024generative,choi2023creator,neate2019empowering,bala2023towards,wu2022ai,fu2024text,durrant2011automics,billewicz2018silly,shelby2024generative,varanasi2023currently,zhang2022storybuddy,li2023understanding,marquez2016embodied,maeng2022designing,khaled2013design,zhang2023towards,louie2022expressive,hodhod2016closing,bako2023user} \\ \hline 
Automatic & \cite{jonsson2022cracking,moruzzi2024user,wang2024farsight,cremaschi2024steampunk,sharma2024generative,fu2024text,verheijden2023collaborative,mirowski2023co,zheng2024charting,suh2021ai,de2024bias,hwang2022too,zhang2022get,weisz2024design,capel2023human,vear2024jess+,lee2024open,chang2024dark,lyu2024preliminary,wallace2022asynchronous,shi2020emog,arakawa2023catalyst,belghith2024testing,mccormack2019silent,park2023importance,lee2024design,wu2022ai,milton2023see,zhang2022storybuddy,khaled2013design,chung2022artistic,liu2022opal,cheon2017configuring,wang2024lave,lawton2023drawing,yuan2022wordcraft,hodhod2016closing,bako2023user} \\ \bottomrule
\end{tabularx}
\end{table}

\begin{table}
\centering
\caption{The agency distribution the papers in our surveys included.}
\label{tbl:distribution}
\begin{tabularx}{\textwidth}{p{0.15\textwidth}p{0.85\textwidth}}
\toprule
\textbf{Agency Distribution} & \textbf{Papers} \\ \midrule 
Locus & \cite{lin2024text,tholander2023design,moruzzi2024user,hoque2024hallmark,sun2024generative,weisz2022better,kamath2024sound,berge2023designing,vear2024jess+,sun2024metawriter,jones2023embodying,cremaschi2024steampunk,gebreegziabher2023patat,sharma2024generative,benjamin2023entoptic,fu2024text,fan2024contextcam,verheijden2023collaborative,liu2024ai,osone2021buncho,louie2020novice,sun2024understanding,mirowski2023co,zheng2024charting,zhou2023beyond,chang2024co,ayobi2023computational,bircanin2021including,li2023want,lin2021engaging,gmeiner2023exploring,kadoma2024role,suh2021ai,newman2024want,zhang2024mathemyths,wang2023treat,trajkova2024exploring,hwang2022too,lukoff2021design,rezk2024agency,praveena2023exploring,han2024teachers,zhang2022get,weisz2024design,wan2024metamorpheus,elagroudy2024transforming,kang2024design,brand2023envisioning,liu2024he,varanasi2024saharaline,lu2022bridging,capel2023human,lee2024open,chang2024dark,lyu2024preliminary,wallace2022asynchronous,shi2020emog,inie2023designing,almeda2024prompting,lazar2018making,belghith2024testing,mccormack2019silent,albaugh2023augmented,reig2023supporting,shibani2024untangling,soler2024arcadia,sharma2023takes,hegemann2023computational,zong2024umwelt,jones2024customization,ashktorab2021effects,lee2024design,von2023designing,leong2024putting,yang2023pair,choi2023crator,neate2019empowering,bala2023towards,wu2022ai,durrant2011automics,billewicz2018silly,shelby2024generative,milton2023see,zhang2022storybuddy,li2023understanding,khaled2013design,chung2022artistic,liu2022opal,dang2022beyond,zhang2023visar,jain2023front,hong2022scholastic,zhang2023towards,cheon2017configuring,wang2024lave,louie2022expressive,lawton2023drawing,yuan2022wordcraft,hodhod2016closing,bako2023user} \\ \hline 
Dynamics & \cite{moruzzi2024user,wang2024farsight,berge2023designing,vear2024jess+,cremaschi2024steampunk,gebreegziabher2023patat,fan2024contextcam,verheijden2023collaborative,liu2024ai,claisse2024understanding,louie2020novice,sun2024understanding,mirowski2023co,zhou2023beyond,chang2024co,bircanin2021including,li2023want,mucha2020co,gmeiner2023exploring,de2024bias,wang2023treat,trajkova2024exploring,lukoff2021design,rezk2024agency,praveena2023exploring,bauer2024musictraces,shaer2024ai,kang2024design,brand2023envisioning,liu2024he,guo2024exploring,lee2024open,chang2024dark,arakawa2023catalyst,wang2024critical,simpson2023rethinking,soler2024arcadia,hegemann2023computational,zong2024umwelt,ashktorab2021effects,chakrabarty2024art,yang2023pair,choi2023creator,neate2019empowering,wu2022ai,durrant2011automics,billewicz2018silly,varanasi2023currently,milton2023see,zhang2022storybuddy,li2023understanding,marquez2016embodied,maeng2022designing,khaled2013design,chung2022artistic,liu2022opal,hong2022scholastic,zhang2023towards,wang2024lave,louie2022expressive,yuan2022wordcraft,hodhod2016closing} \\ \hline 
Granularity & \cite{dhillon2024shaping,sun2024generative,kamath2024sound,berge2023designing,vear2024jess+,gebreegziabher2023patat,fu2024text,fan2024contextcam,suh2024luminate,osone2021buncho,louie2020novice,sun2024understanding,mirowski2023co,zhou2023beyond,ayobi2023computational,bircanin2021including,li2023want,gmeiner2023exploring,suh2021ai,newman2024want,anthraper2024peerconnect,wang2023treat,trajkova2024exploring,hwang2022too,lukoff2021design,praveena2023exploring,han2024teachers,zhang2022get,weisz2024design,brand2023envisioning,lu2022bridging,lee2024open,chang2024dark,wallace2022asynchronous,shi2020emog,inie2023designing,almeda2024prompting,mahdavi2024ai,mccormack2019silent,hegemann2023computational,zong2024umwelt,park2023importance,choi2023creator,neate2019empowering,wu2022ai,durrant2011automics,shelby2024generative,milton2023see,zhang2022storybuddy,li2023understanding,khaled2013design,chung2022artistic,liu2022opal,dang2022beyond,zhang2023visar,jain2023front,hong2022scholastic,zhang2023towards,wang2024lave,louie2022expressive,lawton2023drawing,yuan2022wordcraft,bako2023user} \\ \bottomrule
\end{tabularx}
\end{table}

\begin{table}
\centering
\caption{The control mechanisms the papers in our surveys included.}
\label{tbl:control}
\begin{tabularx}{\textwidth}{p{0.3\textwidth}p{0.7\textwidth}}
\toprule
\textbf{Control Mechanisms} & \textbf{Papers} \\ \midrule 
Guided Input Interaction & \cite{dhillon2024shaping,lin2024text,tholander2023design,moruzzi2024user,hoque2024hallmark,sun2024generative,weisz2022better,kamath2024sound,wang2024farsight,cremaschi2024steampunk,gebreegziabher2023patat,sharma2024generative,benjamin2023entoptic,fu2024text,fan2024contextcam,verheijden2023collaborative,liu2024ai,claisse2024understanding,suh2024luminate,osone2021buncho,louie2020novice,sun2024understanding,mirowski2023co,mirowski2023co,zhou2023beyond,chang2024co,ayobi2023computational,li2023want,lin2021engaging,gmeiner2023exploring,suh2021ai,newman2024want,anthraper2024peerconnect,zhang2024mathemyths,trajkova2024exploring,lukoff2021design,rezk2024agency,praveena2023exploring,han2024teachers,bauer2024musictraces,zhang2022get,weisz2024design,albaugh2023augmented,wan2024metamorpheus,elagroudy2024transforming,kang2024design,brand2023envisioning,liu2024he,lu2022bridging,capel2023human,vear2024jess+,lee2024open,chang2024dark,chang2024dark,lyu2024preliminary,wallace2022asynchronous,shi2020emog,almeda2024prompting,lazar2018making,mahdavi2024ai,arakawa2023catalyst,arakawa2023catalyst} \\ \hline 
Context Awareness and Memory Retention & \cite{tholander2023design,wang2024farsight,gebreegziabher2023patat,sharma2024generative,benjamin2023entoptic,fu2024text,fan2024contextcam,verheijden2023collaborative,liu2024ai,claisse2024understanding,suh2024luminate,osone2021buncho,louie2020novice,sun2024understanding,zheng2024charting,zhou2023beyond,chang2024co,ayobi2023computational,bircanin2021including,li2023want,lin2021engaging,gmeiner2023exploring,kadoma2024role,suh2021ai,newman2024want,anthraper2024peerconnect,zhang2024mathemyths,de2024bias,wang2023treat,trajkova2024exploring,hwang2022too,rezk2024agency,praveena2023exploring,han2024teachers,bauer2024musictraces,zhang2022get,weisz2024design,elagroudy2024transforming,kang2024design,lee2024open,arakawa2023catalyst} \\ \hline 
Transparency and Explainability in Human-AI Agency & \cite{tholander2023design,moruzzi2024user,hoque2024hallmark,sun2024generative,wang2024farsight,berge2023designing,cremaschi2024steampunk,gebreegziabher2023patat,sharma2024generative,benjamin2023entoptic,fu2024text,verheijden2023collaborative,liu2024ai,claisse2024understanding,suh2024luminate,osone2021buncho,louie2020novice,sun2024understanding,mirowski2023co,zheng2024charting,zhou2023beyond,chang2024co,ayobi2023computational,li2023want,lin2021engaging,mucha2020co,gmeiner2023exploring,kadoma2024role,suh2021ai,newman2024want,anthraper2024peerconnect,zhang2024mathemyths,de2024bias,wang2023treat,trajkova2024exploring,hwang2022too,lukoff2021design,rezk2024agency,praveena2023exploring,han2024teachers,zhang2022get,weisz2024design,albaugh2023augmented,wan2024metamorpheus,elagroudy2024transforming,kang2024design,brand2023envisioning,liu2024he,varanasi2024saharaline,guo2024exploring,lu2022bridging,capel2023human,vear2024jess+,lee2024open,chang2024dark,lyu2024preliminary,wallace2022asynchronous,wallace2022asynchronous,shi2020emog,inie2023designing,almeda2024prompting,arakawa2023catalyst} \\ \hline 
Multimodal Action Space Exploration & \cite{tholander2023design,wang2024farsight,berge2023designing,benjamin2023entoptic,fu2024text,fan2024contextcam,verheijden2023collaborative,liu2024ai,claisse2024understanding,suh2024luminate,osone2021buncho,sun2024understanding,zheng2024charting,zhou2023beyond,ayobi2023computational,bircanin2021including,gmeiner2023exploring,newman2024want,anthraper2024peerconnect,zhang2024mathemyths,de2024bias,hwang2022too,praveena2023exploring,bauer2024musictraces,weisz2024design,wan2024metamorpheus,kang2024design,brand2023envisioning,lu2022bridging,capel2023human,vear2024jess+,lee2024open,lyu2024preliminary,wallace2022asynchronous,shi2020emog,almeda2024prompting} \\ \hline
Action Coordination & \cite{dhillon2024shaping,hoque2024hallmark,wang2024farsight,cremaschi2024steampunk,gebreegziabher2023patat,fu2024text,fan2024contextcam,verheijden2023collaborative,liu2024ai,suh2024luminate,osone2021buncho,louie2020novice,mirowski2023co,zheng2024charting,zhou2023beyond,chang2024co,ayobi2023computational,bircanin2021including,li2023want,lin2021engaging,mucha2020co,gmeiner2023exploring,suh2021ai,anthraper2024peerconnect,zhang2024mathemyths,de2024bias,wang2023treat,trajkova2024exploring,hwang2022too,rezk2024agency,praveena2023exploring,han2024teachers,bauer2024musictraces,zhang2022get,weisz2024design,albaugh2023augmented,brand2023envisioning,liu2024he,varanasi2024saharaline,guo2024exploring,capel2023human,vear2024jess+,lee2024open,shi2020emog,inie2023designing,lazar2018making,mahdavi2024ai} \\ \hline 
Attention-focused Processing & \cite{hoque2024hallmark,kamath2024sound,weisz2024design,mahdavi2024ai} \\ \hline  
Confidence Level & \cite{gebreegziabher2023patat,sharma2024generative,zheng2024charting,kadoma2024role,anthraper2024peerconnect,lukoff2021design,weisz2024design} \\ \hline 
Explanatory Feedback Emphasis & \cite{lin2024text,hoque2024hallmark,sun2024metawriter,gebreegziabher2023patat,benjamin2023entoptic,liu2024ai,louie2020novice,ayobi2023computational,newman2024want,de2024bias,trajkova2024exploring,hwang2022too,weisz2024design,elagroudy2024transforming,brand2023envisioning,guo2024exploring,mahdavi2024ai} \\ \hline  
Iterative Feedback Loop & \cite{lin2024text,tholander2023design,moruzzi2024user,hoque2024hallmark,kamath2024sound,wang2024farsight,berge2023designing,cremaschi2024steampunk,gebreegziabher2023patat,sharma2024generative,benjamin2023entoptic,fan2024contextcam,verheijden2023collaborative,liu2024ai,claisse2024understanding,suh2024luminate,osone2021buncho,louie2020novice,sun2024understanding,mirowski2023co,zheng2024charting,zhou2023beyond,chang2024co,ayobi2023computational,bircanin2021including,li2023want,lin2021engaging,mucha2020co,gmeiner2023exploring,kadoma2024role,suh2021ai,newman2024want,anthraper2024peerconnect,zhang2024mathemyths,de2024bias,trajkova2024exploring,hwang2022too,lukoff2021design,rezk2024agency,praveena2023exploring,han2024teachers,bauer2024musictraces,zhang2022get,weisz2024design,albaugh2023augmented,wan2024metamorpheus,elagroudy2024transforming,kang2024design,brand2023envisioning,liu2024he,varanasi2024saharaline,guo2024exploring,lu2022bridging,capel2023human,vear2024jess+,lee2024open,chang2024dark,lyu2024preliminary,wallace2022asynchronous,shi2020emog,inie2023designing,almeda2024prompting,mahdavi2024ai,arakawa2023catalyst} \\ \hline 
Modification and Intervention & \cite{dhillon2024shaping,moruzzi2024user,hoque2024hallmark,weisz2022better,wang2024farsight,berge2023designing,cremaschi2024steampunk,gebreegziabher2023patat,sharma2024generative,benjamin2023entoptic,claisse2024understanding,louie2020novice,sun2024understanding,zhou2023beyond,kadoma2024role,suh2021ai,zhang2024mathemyths,de2024bias,wang2023treat,lukoff2021design,rezk2024agency,praveena2023exploring,han2024teachers,weisz2024design,varanasi2024saharaline,lu2022bridging,capel2023human,vear2024jess+,lee2024open,chang2024dark,lyu2024preliminary,shi2020emog,inie2023designing,almeda2024prompting,arakawa2023catalyst} \\ \hline 
Adaptive Scaffolding & \cite{dhillon2024shaping,berge2023designing,fu2024text,verheijden2023collaborative,liu2024ai,suh2024luminate,osone2021buncho,louie2020novice,chang2024co,ayobi2023computational,li2023want,lin2021engaging,mucha2020co,gmeiner2023exploring,suh2021ai,newman2024want,anthraper2024peerconnect,zhang2024mathemyths,trajkova2024exploring,lukoff2021design,rezk2024agency,praveena2023exploring,han2024teachers,bauer2024musictraces,zhang2022get,weisz2024design,wan2024metamorpheus,brand2023envisioning,liu2024he,varanasi2024saharaline,vear2024jess+,chang2024dark,wallace2022asynchronous,inie2023designing,arakawa2023catalyst} \\ \hline 
Chain of Thought & \cite{fan2024contextcam,liu2024ai,mirowski2023co,zheng2024charting,zhang2022get,weisz2024design} \\
\bottomrule
\end{tabularx}
\end{table}

\begin{table}
\centering
\caption{Applications in the papers in our surveys included.}

\label{tbl:application}
\begin{tabularx}{\textwidth}{p{0.3\textwidth}p{0.7\textwidth}}
\toprule
\textbf{Application} & \textbf{Papers} \\ \midrule 
News Media & \cite{wang2024farsight,sharma2024generative,rezk2024agency,capel2023human,lee2024design,varanasi2023currently,liu2022opal,dang2022beyond}\\ \hline 
Healthcare & \cite{tholander2023design,wang2024farsight,berge2023designing,claisse2024understanding,sun2024understanding,mirowski2023co,chang2024co,bircanin2021including,mucha2020co,bauer2024musictraces,zhang2022get,wan2024metamorpheus,liu2024he,capel2023human,lee2024open,lazar2018making,soler2024arcadia,lee2024design,varanasi2023currently,milton2023see,cheon2017configuring}\\ \hline
Artistic Creation & \cite{dhillon2024shaping,lin2024text,tholander2023design,hoque2024hallmark,kamath2024sound,jones2023embodying,cremaschi2024steampunk,benjamin2023entoptic,fan2024contextcam,suh2024luminate,osone2021buncho,louie2020novice,zheng2024charting,chang2024co,gmeiner2023exploring,suh2021ai,newman2024want,de2024bias,trajkova2024exploring,hwang2022too,praveena2023exploring,han2024teachers,bauer2024musictraces,shaer2024ai,wan2024metamorpheus,kang2024design,brand2023envisioning,guo2024exploring,lu2022bridging,capel2023human,vear2024jess+,chang2024dark,lyu2024preliminary,wallace2022asynchronous,shi2020emog,inie2023designing,almeda2024prompting,mahdavi2024ai,arakawa2023catalyst,mccormack2019silent,jones2023embodying,albaugh2023augmented,wang2024critical,sharma2023takes,hegemann2023computational,lee2024design,von2023designing,chakrabarty2024art,neate2019empowering,billewicz2018silly,shelby2024generative,khaled2013design,chung2022artistic,zhang2023visar,li2023beyond,zhang2023towards,wang2024lave,louie2022expressive,lawton2023drawing,yuan2022wordcraft,hodhod2016closing}\\ \hline 
Education \& Research & \cite{dhillon2024shaping,lin2024text,hoque2024hallmark,weisz2022better,wang2024farsight,sun2024metawriter,gebreegziabher2023patat,benjamin2023entoptic,liu2024ai,zhou2023beyond,ayobi2023computational,li2023want,lin2021engaging,mucha2020co,gmeiner2023exploring,newman2024want,anthraper2024peerconnect,zhang2024mathemyths,wang2023treat,han2024teachers,shaer2024ai,elagroudy2024transforming,varanasi2024saharaline,guo2024exploring,lu2022bridging,capel2023human,lee2024open,chang2024dark,lyu2024preliminary,wallace2022asynchronous,inie2023designing,arakawa2023catalyst,belghith2024testing,albaugh2023augmented,wang2024critical,shibani2024untangling,park2023importance,lee2024design,chakrabarty2024art,leong2024putting,yang2023pair,bala2023towards,wu2022ai,zhang2022storybuddy,li2023understanding,khaled2013design,dang2022beyond,zhang2023visar,hong2022scholastic,li2023beyond,cheon2017configuring,bako2023user} \\ \hline 
Entertainment & \cite{moruzzi2024user,sun2024generative,benjamin2023entoptic,fan2024contextcam,verheijden2023collaborative,suh2024luminate,louie2020novice,zheng2024charting,wang2023treat,trajkova2024exploring,lukoff2021design,praveena2023exploring,capel2023human,lyu2024preliminary,inie2023designing,wang2024critical,reig2023supporting,simpson2023rethinking,sharma2023takes,ashktorab2021effects,choi2023creator,durrant2011automics,marquez2016embodied,khaled2013design,jain2023front,li2023beyond,zhang2023towards,louie2022expressive,hodhod2016closing}\\ \hline
Software Development & \cite{weisz2022better,wang2024farsight,lu2022bridging,capel2023human,lyu2024preliminary,inie2023designing,hegemann2023computational,wu2022ai,varanasi2023currently,li2023beyond,bako2023user}\\ \hline
Accessibility & \cite{fu2024text,chang2024co,kadoma2024role,bauer2024musictraces,lu2022bridging,capel2023human,vear2024jess+,lee2024open,wallace2022asynchronous,lazar2018making,soler2024arcadia,zong2024umwelt,jones2024customization,lee2024design,neate2019empowering,bala2023towards,wu2022ai,li2023understanding,maeng2022designing,jain2023front}\\ \hline
\end{tabularx}
\end{table}

\begin{table}
\centering
\caption{Challenges \& Directions in the papers in our surveys included.}

\label{tbl:application}
\begin{tabularx}{\textwidth}{p{0.3\textwidth}p{0.7\textwidth}}
\toprule
\textbf{Challenges \& Directions} & \textbf{Papers} \\ \midrule 
Creativity and Ownership & \cite{dhillon2024shaping,lin2024text,tholander2023design,moruzzi2024user,hoque2024hallmark,sun2024generative,kamath2024sound,sun2024metawriter,cremaschi2024steampunk,benjamin2023entoptic,fu2024text,fan2024contextcam,verheijden2023collaborative,liu2024ai,suh2024luminate,osone2021buncho,louie2020novice,sun2024understanding,mirowski2023co,zheng2024charting,zhou2023beyond,chang2024co,bircanin2021including,li2023want,gmeiner2023exploring,kadoma2024role,suh2021ai,newman2024want,zhang2024mathemyths,de2024bias,trajkova2024exploring,hwang2022too,praveena2023exploring,han2024teachers,bauer2024musictraces,weisz2024design,shaer2024ai,wan2024metamorpheus,kang2024design,brand2023envisioning,liu2024he,guo2024exploring,lu2022bridging,vear2024jess+,lee2024open,chang2024dark,lyu2024preliminary,wallace2022asynchronous,shi2020emog,inie2023designing,almeda2024prompting,lazar2018making,mahdavi2024ai,belghith2024testing,mccormack2019silent,albaugh2023augmented,wang2024critical,reig2023supporting,shibani2024untangling,simpson2023rethinking,hegemann2023computational,jones2024customization,ashktorab2021effects,park2023importance,lee2024design,chakrabarty2024art,choi2023creator,neate2019empowering,bala2023towards,wu2022ai,durrant2011automics,billewicz2018silly,shelby2024generative,zhang2022storybuddy,li2023understanding,marquez2016embodied,khaled2013design,chung2022artistic,liu2022opal,dang2022beyond,zhang2023visar,zhang2023towards,cheon2017configuring,lawton2023drawing,yuan2022wordcraft,hodhod2016closing,bako2023user}\\ \hline 
Trust and Transparency & \cite{dhillon2024shaping,lin2024text,tholander2023design,moruzzi2024user,hoque2024hallmark,sun2024generative,weisz2022better,kamath2024sound,wang2024farsight,berge2023designing,sun2024metawriter,gebreegziabher2023patat,sharma2024generative,benjamin2023entoptic,fu2024text,fan2024contextcam,verheijden2023collaborative,liu2024ai,claisse2024understanding,suh2024luminate,osone2021buncho,louie2020novice,sun2024understanding,mirowski2023co,zheng2024charting,zhou2023beyond,chang2024co,ayobi2023computational,bircanin2021including,li2023want,lin2021engaging,mucha2020co,gmeiner2023exploring,kadoma2024role,suh2021ai,newman2024want,anthraper2024peerconnect,zhang2024mathemyths,de2024bias,wang2023treat,trajkova2024exploring,hwang2022too,lukoff2021design,rezk2024agency,praveena2023exploring,han2024teachers,bauer2024musictraces,zhang2022get,weisz2024design,shaer2024ai,wan2024metamorpheus,elagroudy2024transforming,kang2024design,brand2023envisioning,liu2024he,varanasi2024saharaline,guo2024exploring,lu2022bridging,capel2023human,vear2024jess+,lee2024open,chang2024dark,lyu2024preliminary,wallace2022asynchronous,inie2023designing,almeda2024prompting,lazar2018making,mahdavi2024ai,arakawa2023catalyst,belghith2024testing,mccormack2019silent,albaugh2023augmented,wang2024critical,reig2023supporting,shibani2024untangling,simpson2023rethinking,soler2024arcadia,sharma2023takes,zong2024umwelt,jones2024customization,ashktorab2021effects,park2023importance,lee2024design,davis2016empirically,chakrabarty2024art,leong2024putting,yang2023pair,choi2023creator,bala2023towards,wu2022ai,durrant2011automics,billewicz2018silly,shelby2024generative,varanasi2023currently,milton2023see,zhang2022storybuddy,li2023understanding,marquez2016embodied,maeng2022designing,khaled2013design,chung2022artistic,liu2022opal,dang2022beyond,zhang2023visar,jain2023front,zhang2023towards,cheon2017configuring,wang2024lave,wang2024lave,louie2022expressive,lawton2023drawing,yuan2022wordcraft,hodhod2016closing,bako2023user}\\ \hline
Data Privacy and Security & \cite{moruzzi2024user,hoque2024hallmark,sun2024generative,weisz2022better,kamath2024sound,wang2024farsight,berge2023designing,gebreegziabher2023patat,benjamin2023entoptic,fu2024text,fan2024contextcam,claisse2024understanding,louie2020novice,sun2024understanding,zheng2024charting,chang2024co,ayobi2023computational,kadoma2024role,anthraper2024peerconnect,zhang2024mathemyths,wang2023treat,hwang2022too,lukoff2021design,rezk2024agency,zhang2022get,weisz2024design,wan2024metamorpheus,elagroudy2024transforming,kang2024design,brand2023envisioning,varanasi2024saharaline,capel2023human,lee2024open,chang2024dark,wallace2022asynchronous,arakawa2023catalyst,soler2024arcadia,park2023importance,lee2024design,davis2016empirically,leong2024putting,yang2023pair,durrant2011automics,billewicz2018silly,varanasi2023currently,milton2023see,maeng2022designing,dang2022beyond,zhang2023visar,jain2023front,louie2022expressive,yuan2022wordcraft}\\ \hline 
Interoperability & \cite{tholander2023design,moruzzi2024user,sun2024generative,weisz2022better,wang2024farsight,berge2023designing,gebreegziabher2023patat,sharma2024generative,verheijden2023collaborative,suh2024luminate,zhou2023beyond,ayobi2023computational,trajkova2024exploring,hwang2022too,praveena2023exploring,zhang2022get,weisz2024design,wan2024metamorpheus,elagroudy2024transforming,liu2024he,guo2024exploring,lu2022bridging,capel2023human,vear2024jess+,arakawa2023catalyst,sharma2023takes,hegemann2023computational,zong2024umwelt,jones2024customization,ashktorab2021effects,lee2024design,wu2022ai,durrant2011automics,zhang2022storybuddy,chung2022artistic,liu2022opal,dang2022beyond,zhang2023visar,jain2023front,wang2024lave,yuan2022wordcraft} \\ \hline 
Social Impact and Equity & \cite{dhillon2024shaping,tholander2023design,moruzzi2024user,hoque2024hallmark,sun2024generative,kamath2024sound,wang2024farsight,berge2023designing,sun2024metawriter,cremaschi2024steampunk,gebreegziabher2023patat,sharma2024generative,benjamin2023entoptic,fu2024text,fan2024contextcam,liu2024ai,claisse2024understanding,louie2020novice,mirowski2023co,zheng2024charting,zhou2023beyond,chang2024co,bircanin2021including,li2023want,lin2021engaging,mucha2020co,kadoma2024role,suh2021ai,newman2024want,anthraper2024peerconnect,de2024bias,wang2023treat,hwang2022too,lukoff2021design,rezk2024agency,praveena2023exploring,bauer2024musictraces,weisz2024design,brand2023envisioning,varanasi2024saharaline,capel2023human,chang2024dark,lyu2024preliminary,lazar2018making,sharma2023takes,davis2016empirically,choi2023creator,neate2019empowering,bala2023towards,shelby2024generative,varanasi2023currently,milton2023see,cheon2017configuring}\\ \hline
Ethical and Bias Concerns & \cite{dhillon2024shaping,lin2024text,tholander2023design,moruzzi2024user,hoque2024hallmark,sun2024generative,weisz2022better,kamath2024sound,wang2024farsight,berge2023designing,sun2024metawriter,gebreegziabher2023patat,sharma2024generative,fu2024text,verheijden2023collaborative,liu2024ai,suh2024luminate,osone2021buncho,sun2024understanding,mirowski2023co,zheng2024charting,zhou2023beyond,chang2024co,ayobi2023computational,li2023want,lin2021engaging,mucha2020co,gmeiner2023exploring,kadoma2024role,suh2021ai,newman2024want,zhang2024mathemyths,de2024bias,wang2023treat,trajkova2024exploring,hwang2022too,lukoff2021design,rezk2024agency,han2024teachers,zhang2022get,weisz2024design,shaer2024ai,wan2024metamorpheus,elagroudy2024transforming,kang2024design,brand2023envisioning,liu2024he,varanasi2024saharaline,guo2024exploring,lu2022bridging,capel2023human,vear2024jess+,lee2024open,chang2024dark,lyu2024preliminary,wallace2022asynchronous,inie2023designing,almeda2024prompting,lazar2018making,mahdavi2024ai,arakawa2023catalyst,belghith2024testing,mccormack2019silent,wang2024critical,reig2023supporting,shibani2024untangling,simpson2023rethinking,soler2024arcadia,sharma2023takes,hegemann2023computational,zong2024umwelt,ashktorab2021effects,park2023importance,lee2024design,davis2016empirically,chakrabarty2024art,leong2024putting,yang2023pair,choi2023creator,neate2019empowering,bala2023towards,wu2022ai,durrant2011automics,billewicz2018silly,shelby2024generative,varanasi2023currently,milton2023see,zhang2022storybuddy,li2023understanding,marquez2016embodied,maeng2022designing,khaled2013design,chung2022artistic,liu2022opal,dang2022beyond,zhang2023visar,jain2023front,cheon2017configuring,wang2024lave,louie2022expressive,lawton2023drawing,yuan2022wordcraft,hodhod2016closing,bako2023user}\\ \hline
Long-Term Paradigm Shift & \cite{wang2024farsight,sun2024metawriter,cremaschi2024steampunk,fu2024text,liu2024ai,shaer2024ai,elagroudy2024transforming,guo2024exploring,chang2024dark,lee2024design,khaled2013design,yuan2022wordcraft}\\ \hline
\end{tabularx}
\end{table}

\end{document}